\documentclass[prl,nofootinbib,superscriptaddress,twocolumn]{revtex4}
\def\mysection#1{{\bf #1.} }
\def\mysections#1{{\bf #1.} }

\usepackage{amssymb}
\usepackage{amsmath}
\usepackage{graphicx}
\usepackage{longtable}
\usepackage{verbatim}
\usepackage{amsfonts}
\usepackage{color}
\usepackage[bookmarksdepth=4]{hyperref}

\newcommand{\be}{\begin{eqnarray}}
\newcommand{\ee}{\end{eqnarray}}
\newcommand{\bea}{\begin{eqnarray}}
\newcommand{\eea}{\end{eqnarray}}
\newcommand{\beq}{\begin{eqnarray}}
\newcommand{\eeq}{\end{eqnarray}}
\def\beqa{\begin{eqnarray}}
\def\eeqa{\end{eqnarray}}
\newcommand{\no}{\nonumber}
\def\lsim{\mathrel{\rlap{\lower4pt\hbox{\hskip1pt$\sim$}}
    \raise1pt\hbox{$<$}}}         
\def\gsim{\mathrel{\rlap{\lower4pt\hbox{\hskip1pt$\sim$}}
    \raise1pt\hbox{$>$}}}         

\def\ah{\overline{\rm ^3He}}
\def\at{\overline{\rm t}}
\def\R{\mathcal{R}}
\def\X{X_{\rm esc}}

\begin{document}

\vspace*{-30mm}

\title{\boldmath Cosmic rays, anti-helium, and an old navy spotlight}

\author{Kfir Blum}\email{kfir.blum@weizmann.ac.il}
\author{Kenny Chun Yu Ng}\email{chun-yu.ng@weizmann.ac.il}
\author{Ryosuke Sato}\email{ryosuke.sato@weizmann.ac.il}
\affiliation{
Dept.~of Part.~Phys.~\& Astrophys., Weizmann Institute of Science, POB 26, Rehovot, Israel}
\author{Masahiro Takimoto}\email{masahiro.takimoto@weizmann.ac.il}
\affiliation{
Dept.~of Part.~Phys.~\& Astrophys., Weizmann Institute of Science, POB 26, Rehovot, Israel}
\affiliation{
Theory Center, KEK, 1-1 Oho, Tsukuba, Ibaraki 305-0801, Japan}
\vspace*{1cm}

\begin{abstract}
\noindent
Cosmic-ray anti-deuterium and anti-helium have long been suggested as probes of dark matter, as their secondary astrophysical production was thought extremely scarce. But how does one actually predict the secondary flux? Anti-nuclei are dominantly produced in pp collisions, where laboratory cross section data is lacking. We make a new attempt at tackling this problem by appealing to a scaling law of nuclear coalescence with the physical volume of the hadronic emission region. The same volume is probed by Hanbury Brown-Twiss (HBT) two-particle correlations. We demonstrate the consistency of the scaling law with systems ranging from central and off-axis AA collisions to pA collisions, spanning 3 orders of magnitude in coalescence yield. Extending the volume scaling to the pp system, HBT data allows us to make a new estimate of coalescence, that we test against preliminary ALICE pp data.  For anti-helium the resulting cross section is 1-2 orders of magnitude higher than most earlier estimates. The astrophysical secondary flux of anti-helium could be within reach of a five-year exposure of AMS02.
 \end{abstract}

\maketitle

\noindent
\mysection{Introduction}
\noindent
Composite cosmic-ray (CR) anti-nuclei like anti-deuterium ($\bar d$) and anti-helium ($\ah$) have long been suggested as probes of dark matter~\cite{Donato:1999gy, Baer:2005tw, Donato:2008yx, Brauninger:2009pe, Kadastik:2009ts, Cui:2010ud, Dal:2012my, Ibarra:2012cc, Fornengo:2013osa, Carlson:2014ssa, Aramaki:2015pii}, as their secondary astrophysical production was thought to be negligible~\cite{Chardonnet:1997dv,Duperray:2005si,Ibarra:2013qt,Cirelli:2014qia,Herms:2016vop}. But how does one actually predict the secondary flux? Astrophysical anti-nuclei are dominantly produced in pp collisions, for which laboratory cross section data is scarce or altogether absent. 

We make a new attempt at tackling this problem by appealing to a scaling law of nuclear coalescence with the physical volume of the hadronic emission region. The same volume is probed by Hanbury Brown-Twiss (HBT) two-particle correlation measurements~\cite{Scheibl:1998tk,Lisa:2005dd}. A common tool in heavy ion collision studies~\cite{Uribe:1993tr,Malinina:2013fhp,Boggild:1998dx,Adam:2015vja,Bearden:2001sy,Chajecki:2005zw}, the HBT method owes its acronym to the inventors of intensity interferometry, utilised in the 50's for the first angular size determination of a star outside the solar system using two US navy spotlights as light buckets~\cite{Brown:1956zza,HanburyBrown:1956bqd}. In this paper we redirect the HBT idea back to astrophysics, this time in connecting accelerator data to anti-nuclei CRs.

We show that the scaling law applies to systems ranging from central and off-axis AA collisions to pA and pp collisions, spanning 3 orders of magnitude in coalescence yield. Guided by HBT data we make a new estimate of the $pp\to\ah$ cross section, that we validate against preliminary ALICE pp data. 

\begin{figure}[t]
\begin{center}
\includegraphics[scale=0.35]{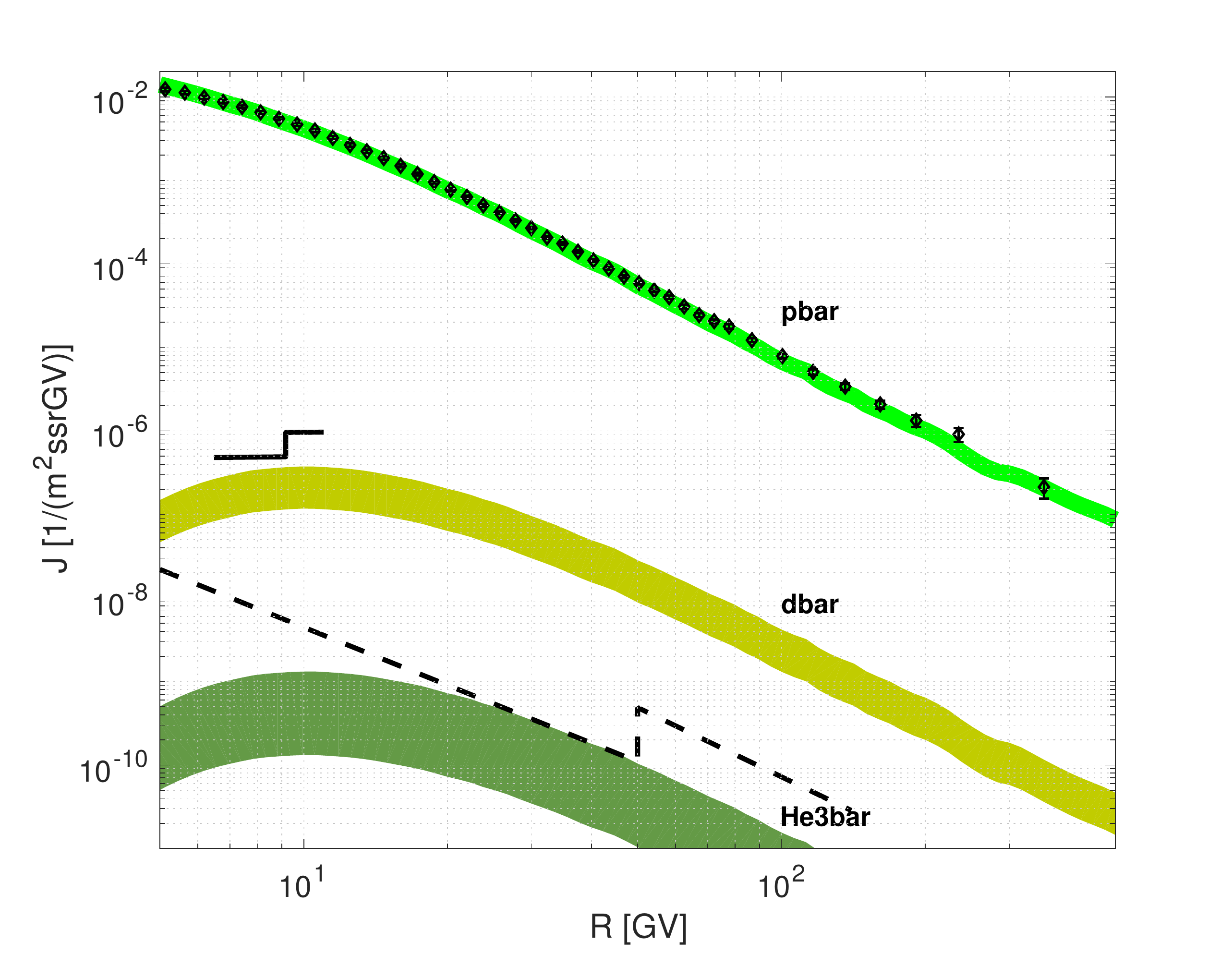}
\caption{Predicted flux of $\bar p,\,\bar d,\,\ah$. AMS02 $\bar p$ data is taken from Ref.~\cite{Aguilar:2016kjl}. AMS02 $\bar d$ flux sensitivity (5-yr, 95\%CL) in the kinetic energy range 2.5-4.7~GeV/nuc, as estimated in Ref.~\cite{Aramaki:2015pii}, is shown in solid line. AMS02 $\ah$ flux sensitivity (5-yr, 95\%CL), derived from the $\ah/$He estimate of Ref.~\cite{kounineHebar}, is shown in dashed line.}
\label{fig:flux}
\end{center}
\end{figure}

Our results for the $\bar p,\,\bar d,$ and $\ah$ flux are summarised in Fig.~\ref{fig:flux}. The predicted $\ah$ yield is 1-2 orders of magnitude higher than most earlier estimates~\cite{Chardonnet:1997dv,Duperray:2005si,Ibarra:2013qt,Cirelli:2014qia,Herms:2016vop} and the flux could reach, within uncertainties, the expected 5-yr 95\%CL flux upper limit estimated for AMS02 prior to its launch~\cite{kounineHebar}.

The outline of the paper is as follows. We begin by calculating the secondary $\bar p$ flux, demonstrating along the way that the astrophysical details of CR propagation are irrelevant for the calculation of stable, relativistic, secondary CR (anti-)nuclei like $\bar p,\,\bar d$ and $\ah$. The challenge, instead, is in computing the production cross sections. Invoking the HBT-coalescence relation, we derive new estimates for the $\bar d$ and $\ah$ yield in pp collisions, forming the basis of our results in Fig.~\ref{fig:flux}. Many details are reserved to the Appendices: A. accelerator data analysis; B. comparison to previous work; C. phase space calculations; D. comments on the $\bar pp$ secondary source; E. benchmark fragmentation cross sections.
\\

\noindent
\mysection{CR $\bar p$: the Galaxy is a fixed-target experiment}
\noindent
CR $\bar p,\,\bar d$ and $\ah$ are produced as secondaries in collisions of other CRs, notably protons, with interstellar matter (ISM), notably hydrogen in the Galaxy. While the details of CR propagation are unknown, the confinement in the Galaxy is magnetic and thus different CR particles that share a common distribution of sources exhibit similar propagation if sampled at the same magnetic rigidity $\R=p/Z$. 
It is therefore natural to gauge the propagation of CR anti-nuclei from that of secondary nuclei like boron (B), formed by fragmentation of heavier CRs. 
For such secondaries, the ratio of densities of two specie $a,b$ satisfies an approximate empirical relation, valid at relativistic energies ($\R\gtrsim$ few GV)~\cite{Engelmann:1990zz,2003ApJ...599..582W,Katz:2009yd},
\be\label{eq:sec}\frac{n_a(\R)}{n_b(\R)}&=&\frac{Q_a(\R)}{Q_b(\R)}.\ee
Here $Q_a$ denotes the net production of species $a$ per unit ISM column density, 
\be\label{eq:Q} Q_a(\R)&=&\sum_Pn_P(\R)\frac{\sigma_{P\to a}(\R)}{m}-n_a(\R)\frac{\sigma_a(\R)}{m},\ee
where $\sigma_a/m$ and $\sigma_{P\to a}/m$ are the total inelastic and the partial $P\to a$ cross section per target ISM particle mass $m$, respectively. 
These cross sections can (and for $\bar p$, $\bar d$ and $\ah$, do) depend on energy. In Eq.~(\ref{eq:Q}) we define these cross sections such that the source term $Q_a(\R)$ is proportional to the progenitor species density $n_P(\R)$ expressed at the same rigidity.

Eq.~(\ref{eq:sec}) is theoretically natural, in that it is guaranteed to apply if the relative composition of the CRs (not CR intensity, nor target ISM density) in the regions that dominate the spallation is similar to that measured locally at the solar system~\cite{Ginzburg:1990sk,Katz:2009yd}, and as long as no significant energy gain/loss occurs during propagation. Restricting our analysis to $\R\geq5$~GV, we expect Eq.~(\ref{eq:sec}) to be accurate to order 10\% or so, as demonstrated by  nuclei data~\cite{Engelmann:1990zz,2003ApJ...599..582W,Katz:2009yd,Blum:2013zsa}. 

Eq.~(\ref{eq:sec}) is useful because we can use the measured flux of B, C,  O, p, He,... to predict, e.g., the $\bar p$ flux~\cite{1992ApJ...394..174G,Katz:2009yd,Blum:2013zsa}:
\be\label{eq:pbfromB} n_{\bar p}(\R)&=&\frac{n_{\rm B}(\R)}{Q_{\rm B}(\R)}Q_{\bar p}(\R).\ee
The RHS of Eq.~(\ref{eq:pbfromB}) is derived from laboratory cross section data and from direct measurements of local CRs, without reference to any detail of propagation.

The quantity 
\be\label{eq:Xbc} \X(\R)&=&\frac{n_{\rm B}(\R)}{Q_{\rm B}(\R)},\ee
known as the CR grammage, measures the column density of ISM traversed by CRs. We combine AMS02 B/C~\cite{Aguilar:2016vqr} and C/O~\cite{AMS02:C2O} 
with heavier CR data from HEAO3~\cite{Engelmann:1990zz} and with laboratory fragmentation cross section data (see e.g.~\cite{Tomassetti:2015nha}), to derive $\X$ directly from measurements:
\be\label{eq:X}\X&=&\frac{\rm (B/C)}{\sum_{\rm P=C,N,O,...}{\rm(P/C)}\frac{\sigma_{\rm P\to B}}{m}-{\rm(B/C)}\frac{\sigma_{\rm B}}{m}}.
\ee
Our result for $\X$ agrees with the power-law approximation derived in Ref.~\cite{Blum:2013zsa} to 20\% accuracy.

\begin{figure}[t]
\begin{center}
\includegraphics[scale=0.4]{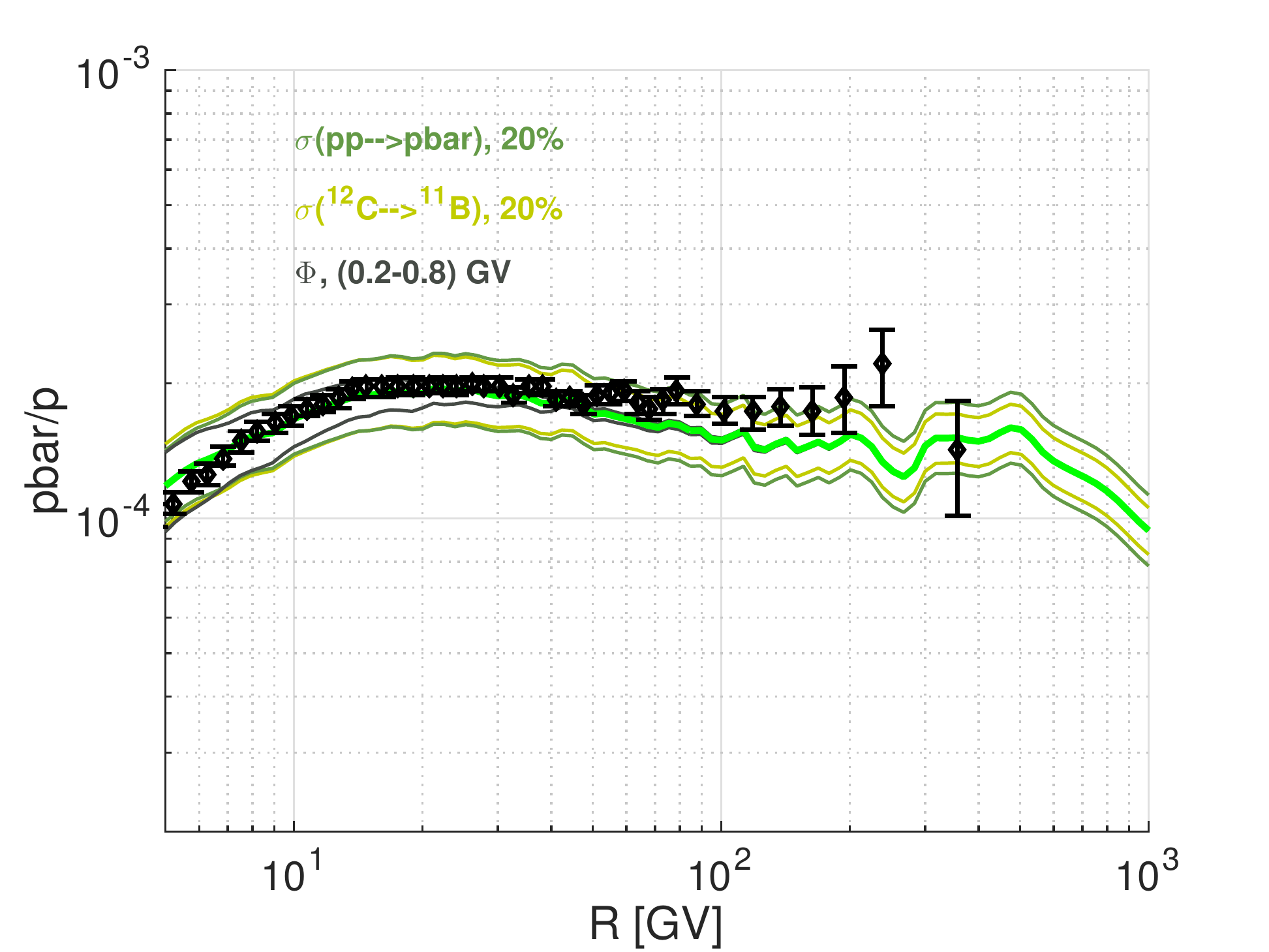}
\caption{Observed $\bar p/p$ ratio~\cite{Aguilar:2016kjl} vs. the secondary prediction. Wiggles in the theory curve come from our direct data-driven use of the CR grammage, and reflect fluctuations in the AMS02 B/C data~\cite{Aguilar:2016vqr}. Thick line is the secondary prediction with input cross sections detailed in App.~E, while thin lines show the response of the prediction for variation in (i) $pp\to\bar p$ cross section within $\pm20\%$, (ii) fragmentation cross section p$^{12}$C$\to$$^{11}$B within $\pm20\%$, (iii) variation in the solar modulation parameter in the range $\Phi=(0.2-0.8)$~GV.}
\label{fig:pppb}
\end{center}
\end{figure}

Now that we have $\X$, we use the $\bar p$ production and loss cross sections parametrised in~\cite{1983JPhG....9..227T,0305-4616-9-10-015} (applying the correction in~\cite{Winkler:2017xor}) together with measurements of the proton and helium~\cite{Aguilar:2015ooa,Aguilar:2015ctt} flux to calculate $Q_{\bar p}$ and apply it in Eq.~(\ref{eq:pbfromB}). Solar modulation is included as in~\cite{2003ApJ...599..582W} with $\Phi=450$~MV. The result is compared to data in Figs.~\ref{fig:flux}-\ref{fig:pppb}. 

Figs.~\ref{fig:flux}-\ref{fig:pppb} demonstrate that the $\bar p$ flux measured by AMS02~\cite{Aguilar:2016kjl} is consistent with secondary production~\cite{Blum:2013zsa}. Beyond this fact, they also demonstrate that -- as far as relativistic, stable, secondary nuclei and anti-nuclei CRs are considered -- the Galaxy is essentially a fixed-target experiment. 
Having calibrated the set-up on one species (B), one can calculate the flux of other secondaries directly from particle physics cross sections. The problem of predicting the anti-nuclei CR flux is therefore decoupled from the modelling of propagation and is reduced to calculating the relevant cross sections, to which we attend next. \\

\noindent
\mysection{Calibrating coalescence with HBT correlations}\label{sec:hbt}
\noindent
We use a coalescence ansatz~\cite{Butler:1963pp,Schwarzschild:1963zz,Gutbrod:1988gt} relating the formation of composite nucleus product with mass number $A$ to the formation cross section of the nucleon constituents:
\be\label{eq:coal}E_A\frac{dN_{A}}{d^3p_{A}}&=&B_A\,R(x)\,\left(E_{p}\frac{dN_{p}}{d^3p_{p}}\right)^A,\ee
where $dN_i=d\sigma_i/\sigma$ is the differential yield, $\sigma$ is the total inelastic cross section, and the constituent momenta are taken at $p_p=p_A/A$.  

The factor $R(x)$, with $x=\sqrt{s+A^2m_p^2-2\sqrt{s}\tilde E_A}$ and $\tilde E_A$ the centre of mass product nucleus energy, is a phase space correction that we define as in~\cite{Duperray:2003tv}. This becomes necessary in order to extend the coalescence analysis down to near-threshold collision energies, important for the astrophysics as well as for low energy laboratory data. Details on the derivation of $R(x)$ are given in App.~C.

$B_A$, the coalescence factor, needs to be extracted from accelerator data.
However, experimental information on $\bar d$ and $\ah$ production is scarce and, in the most part, limited to AA or pA collisions. For pp collisions, the most relevant system for CR astrophysics, no quantitative data exists 
for $pp\to\ah$, and the data for $pp\to\bar d$ is sparse.

Faced with this problem, previous estimates~\cite{Chardonnet:1997dv,Duperray:2005si,Ibarra:2013qt,Cirelli:2014qia,Herms:2016vop} of the secondary CR $\bar d$ and $\ah$ flux made two key simplifying assumptions: 
\begin{enumerate}
\item Coalescence parameters used to fit $pp\to\bar d$ data were translated directly to $pp\to\ah$. More precisely, the coalescence factor $B_A$ was converted to a coalescence momentum $p_c$, via
\be\label{eq:pcB} 
\frac{A}{m_p^{A-1}}\left(\frac{4\pi}{3}p_c^3\right)^{A-1}&=&B_A.\ee
The value of $p_c$ found from $pp\to\bar d$ accelerator data was then assumed to describe $pp\to\ah$.
\item The same coalescence momentum was sometimes assumed to describe both $pA\to\bar d$ and $pp\to\bar d$.
\end{enumerate}
In what follows we give theoretical and empirical evidence, suggesting that both assumptions may be incorrect. To do this, we make an excursion into the physics of coalescence. 

The role of the factor $B_A$ is to capture the probability for A nucleons produced in a collision to merge into a composite nucleus. It is natural for the merger probability to scale as~\cite{Bond:1977fd,Mekjian:1977ei,Csernai:1986qf} 
\be\label{eq:vscale}B_A\propto V^{1-A},\ee
where $V$ is the characteristic volume of the hadronic emission region. 
A model of coalescence that realises the scaling of Eq.~(\ref{eq:vscale}) was presented in Ref.~\cite{Scheibl:1998tk}. 
A key observation in~\cite{Scheibl:1998tk} is that the same hadronic emission volume is probed by Hanbury Brown-Twiss (HBT) two-particle correlation measurements~\cite{Lisa:2005dd}. Both HBT data and nuclear yield measurements are available for  AA and pA systems, allowing a test of  Eq.~(\ref{eq:vscale}). 

Ref.~\cite{Scheibl:1998tk} proposed the following formula for the coalescence factor,
\be\label{eq:BR0}
B_3&=&\frac{\left(2\pi\right)^3}{4\sqrt{3}}\left\langle C_3\right\rangle\left(m_t\,R_1\,R_2\,R_3\right)^{-2}.
\ee
Here, $m_t$ is the transverse mass and $R_i$ are the $m_t$-dependent HBT scales characterising the collision. For concreteness we focus on $A=3$, but the treatment of $A=2$ is analogous. 
The quantity $\left\langle C_3\right\rangle$ expresses the finite support of the $^3$He wave function. It may be estimated via
\be\left\langle C_3\right\rangle&\approx&\Pi_{i=1,2,3}\left(1+\frac{b_3^2}{4R_i^2}\right)^{-1},\ee
where $b_3\approx1.75$~fm is the $^3$He nucleus size.
For $p_t\lesssim1$~GeV, setting $R_i\approx R$, we have
\be\label{eq:BR}
\frac{B_3}{\rm GeV^4}&\approx&0.0024\left(\left(\frac{R(p_t)}{\rm 1~fm}\right)^2+0.8\left(\frac{b_3}{\rm 1.75~fm}\right)^2\right)^{-3}
\ee
The extension to deuterium, with nucleus size $b_2=3.2$~fm, is given by
\be\label{eq:BR2}
\frac{B_2}{\rm GeV^2}&\approx&0.068\left(\left(\frac{R(p_t)}{\rm 1~fm}\right)^2+2.6\left(\frac{b_2}{\rm 3.2~fm}\right)^2\right)^{-\frac{3}{2}}
\ee

\begin{figure}[t]
\begin{center}
\includegraphics[scale=0.55]{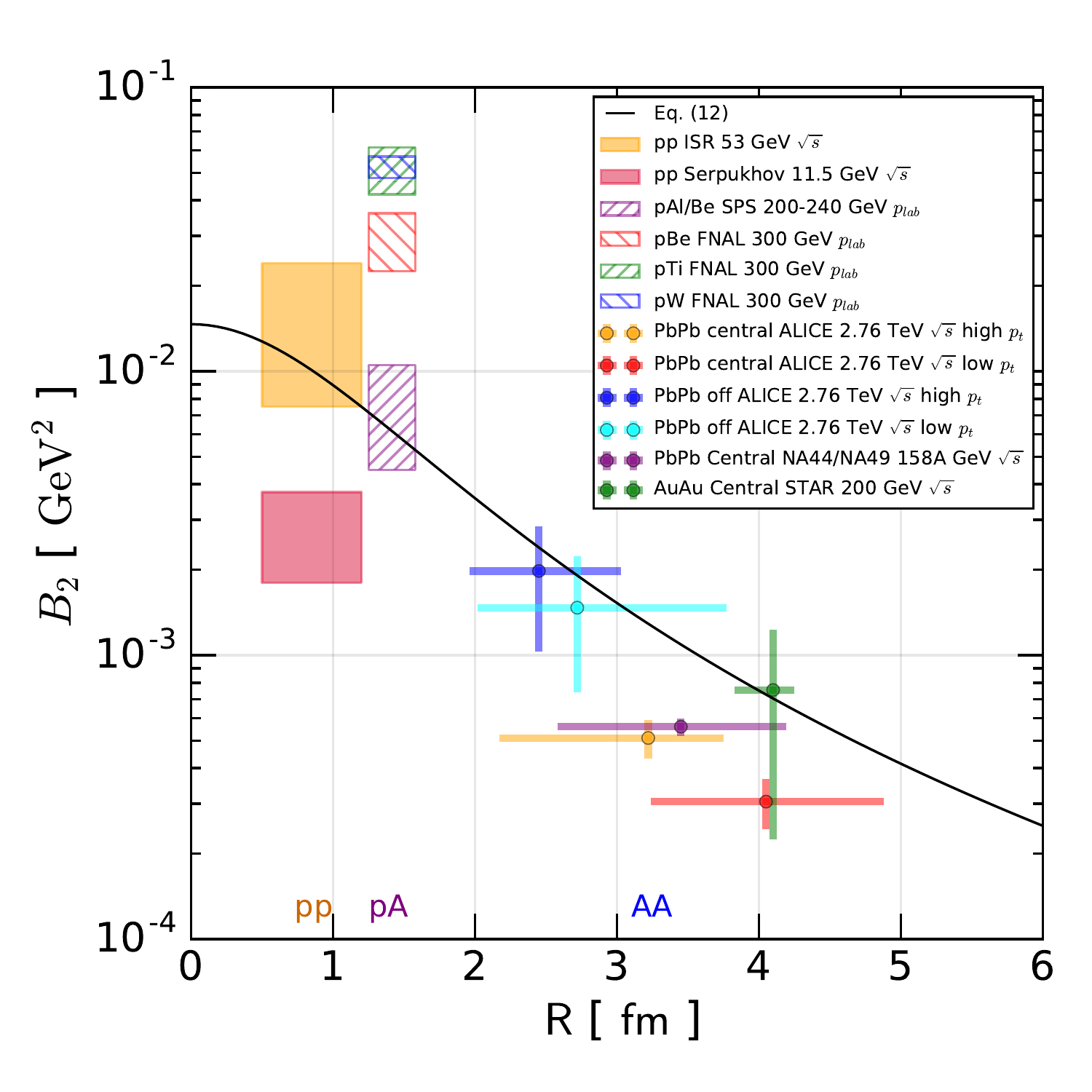}
\includegraphics[scale=0.55]{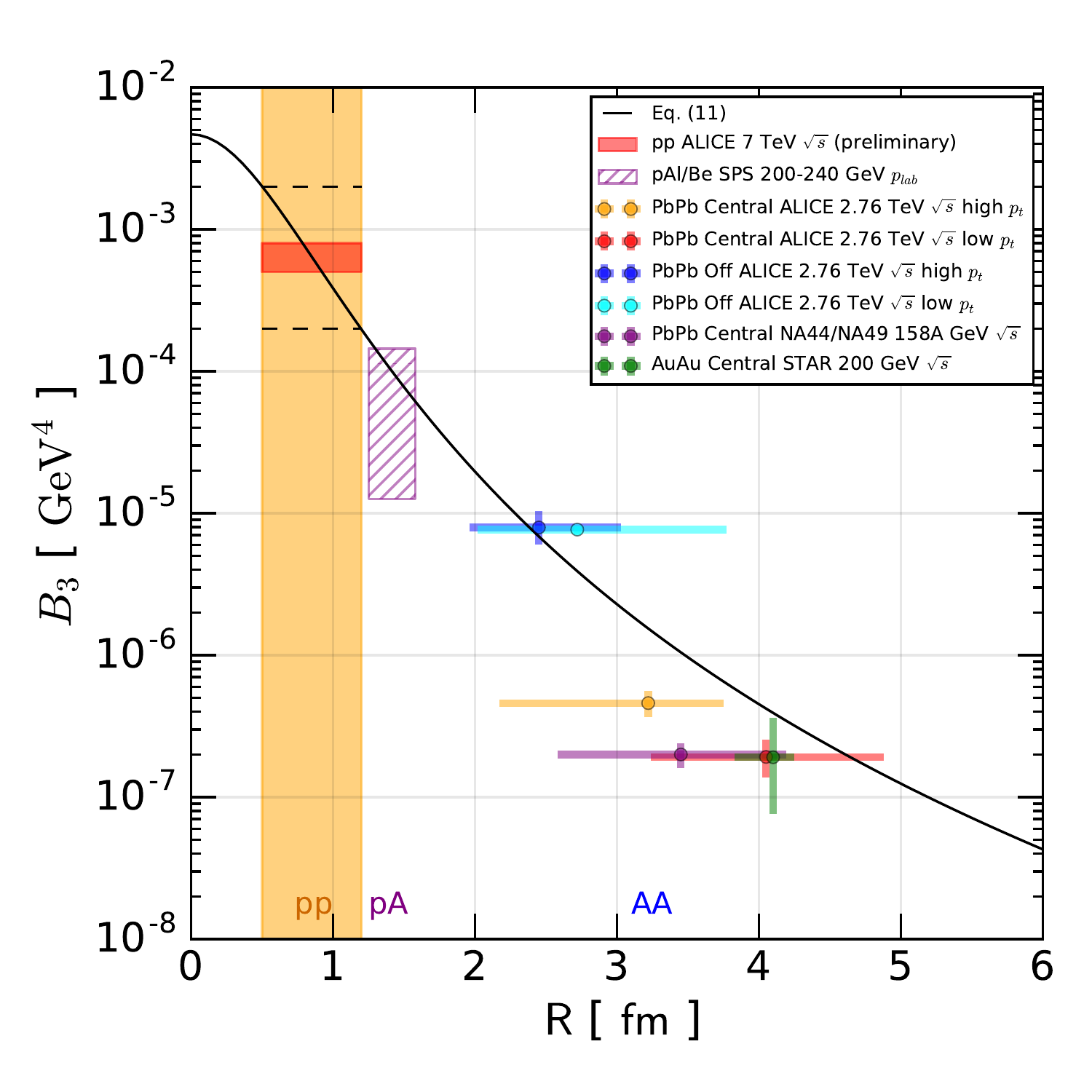}
\caption{Coalescence factor $B_2$ (Top) and $B_3$ (Bottom) vs. HBT radius. The prediction of Eqs.~(\ref{eq:BR}-\ref{eq:BR2}) is shown as solid line. Details of the data analysis are given in App.~A. (Boxes denote systems for which the coalescence factor and the HBT radius are taken from different data sets.)}
\label{fig:Bformula}
\end{center}
\end{figure}

The coalescence factor in AA, pA, and pp collisions, presented w.r.t. HBT scale deduced for the same systems, is shown in Fig.~\ref{fig:Bformula}. The data analysis entering into making the plot is summarised in App.~A. 
The data is consistent with Eqs.~(\ref{eq:BR}) (bottom panel) and~(\ref{eq:BR2}) (top panel), albeit with large uncertainty. 

HBT data for pp collisions~\cite{Uribe:1993tr,Malinina:2013fhp,Aamodt:2010jj} suggest $R$ in the range $0.5-1.2$~fm, indicated by letters in both panels of Fig.~\ref{fig:Bformula}. 
For $pp\to\bar d$, direct measurements from the ISR~\cite{ALBROW1975189,1973PhLB...46..265A,Henning:1977mt} give
\be\label{eq:B2val} B_2^{(pp)}&=&(0.75-2.4)\times10^{-2}~{\rm GeV^2}.
\ee
As seen in the top panel of Fig.~\ref{fig:Bformula}, this result is consistent with the intersect of Eq.~(\ref{eq:BR2}) with the specified range of $R$. (As done in Refs.~\cite{Duperray:2005si,Ibarra:2013qt,Cirelli:2014qia}, we discard here the high-$p_t$ data from Serpukhov~\cite{Abramov:1986ti} and only show it in Fig.~\ref{fig:Bformula} for completeness. Details can be found in App.~A.)

For $pp\to\ah$ we do not have direct experimental information. We therefore extract a rough prediction of $B_3$, by taking the intersect of Eq.~(\ref{eq:BR}) with the two ends of the relevant range for $R$. This gives the following order of magnitude estimate:
\be\label{eq:B3val} B_3^{(pp)}&=&(2-20)\times10^{-4}~{\rm GeV^4}\;\;\,{\rm (HBT-\,based)},\ee
marked by the two horizontal dashed lines in the bottom panel of Fig.~\ref{fig:Bformula}. 

\begin{figure}[t]
\begin{center}
\includegraphics[scale=0.7]{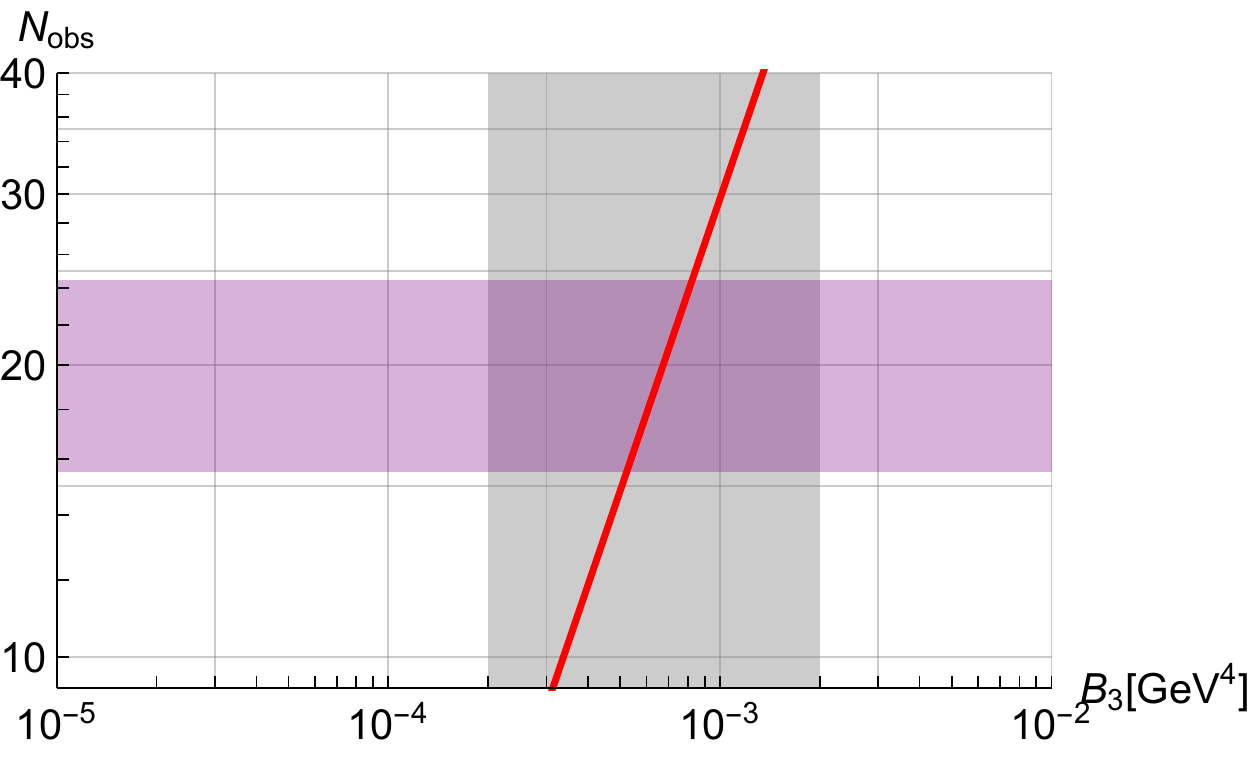}
\caption{Analysis of ALICE pp data~\cite{Sharma:2011ya,Adam:2015vda}. The  number of observed $\ah$ events is shown as horizontal band with 1$\sigma$ Poisson range. Eq.~(\ref{eq:B3val}) is shown by the vertical band. The expected number of events as function of $B_3$, imposing the kinematical cuts and efficiency reported for the analysis, is shown by the red line.}
\label{fig:alice}
\end{center}
\end{figure}

Results from the ALICE experiment allow us to make a preliminary test of Eq.~(\ref{eq:B3val}). Ref.~\cite{Sharma:2011ya} reported 20 $\ah$ and 20 $\at$ in the ALICE pp $\sqrt{s}=7$~TeV run, corresponding to luminosity $\mathcal{L}\approx2.2$~nb$^{-1}$ with a pseudo-rapidity cut $|\eta|<0.9$ and with no further $p_t$ cut\footnote{We thank Natasha Sharma for clarifying the experimental procedure.}. The $p_t$-dependent efficiency for $\ah$ detection was given in~\cite{Adam:2015vda}. In Fig.~\ref{fig:alice} we use these parameters to calculate the expected number of $\ah$ or $\at$ events and compare with data. The result supports a coalescence factor $B^{(pp)}_3\approx(5-8)\times10^{-4}$~GeV$^4$, in agreement with Eq.~(\ref{eq:B3val}). A dedicated analysis by the ALICE collaboration is highly motivated.\\
%

\noindent
\mysection{CR anti-helium}\label{sec:pp}
\noindent
Two channels produce a final state $\ah$: direct $pp\to\ah$ and $pp\to\at$ with subsequent decay $\at\to\ah$. The first channel should suffer some Coulomb suppression with a Gamow factor that can be estimated by $f_{\rm coul}\sim e^{-\frac{\pi\alpha m_p}{p_c}}$. 
Eq.~(\ref{eq:B3val}) suggests $p_c\sim0.1-0.2$~GeV, leading to $f_{\rm coul}\sim0.8-0.9$. This is supported by experimental results on the relative yield $^3$He/t~\cite{Sharma:2011ya,Sharma:2016vpz,Adam:2015vda} that are consistent with $f_{\rm coul}\sim1$. (Ref.~\cite{BUSSIERE19801} reported $\ah/\at<1$; however, the $^3$He/t data from the same publication show an opposite trend, $^3$He/t $\geq1$.) 
In what follows, for concreteness we focus on $pp\to\at$ but we include a factor of 2 increased yield from the direct $pp\to\ah$ channel. 

Combining Eq.~(\ref{eq:B3val}) with the $pp\to\bar p$ production cross section\footnote{A 19\% hyperon contribution to the $\bar p$ cross section~\cite{Blum:2017iol} is subtracted, assuming that coalescence feeds only on prompt $\bar p$ and neglecting the contribution from $\overline{^3_\Lambda {\rm H}}$ decay.}~\cite{1983JPhG....9..227T,0305-4616-9-10-015}, we use  Eq.~(\ref{eq:coal}) to obtain the differential cross section $E_{\at}\frac{d\sigma_{pp\to\at}}{d^3p_{\at}}=\sigma_{pp}E_{\at}\frac{dN_{\at}}{d^3p_{\at}}$, where $\sigma_{pp}$ is the total inelastic pp cross section~\cite{Olive:2016xmw,Fagundes:2012rr}. The effective production cross section to be used in Eq.~(\ref{eq:Q}) is then
\be\label{eq:spp} \sigma_{pp\to\ah}(\R)&=&2\int_{\epsilon}^\infty d\epsilon_p\frac{n_p(\epsilon_p)}{n_p(\R)}\frac{d\sigma_{pp\to\at}(\epsilon_p,\epsilon)}{d\epsilon},
\ee
where
\be\frac{d\sigma_{pp\to\at}(\epsilon_p,\epsilon_{\at})}{d\epsilon_{\at}}&=&2\pi p_{\at}\int dc_\theta \left(\epsilon_{\at}\frac{d\sigma_{pp\to\at}}{d^3p_{\at}}\right).\ee
The final state rigidity and energy are related by $4\R^2=\epsilon^2-9m_p^2$. 
For the inelastic cross section of $\ah$, entering the loss term in Eq.~(\ref{eq:Q}), we use the $\bar p$ cross section~\cite{1983JPhG....9..227T} multiplied by 3. 

\begin{figure}[t]
\begin{center}
\includegraphics[scale=0.45]{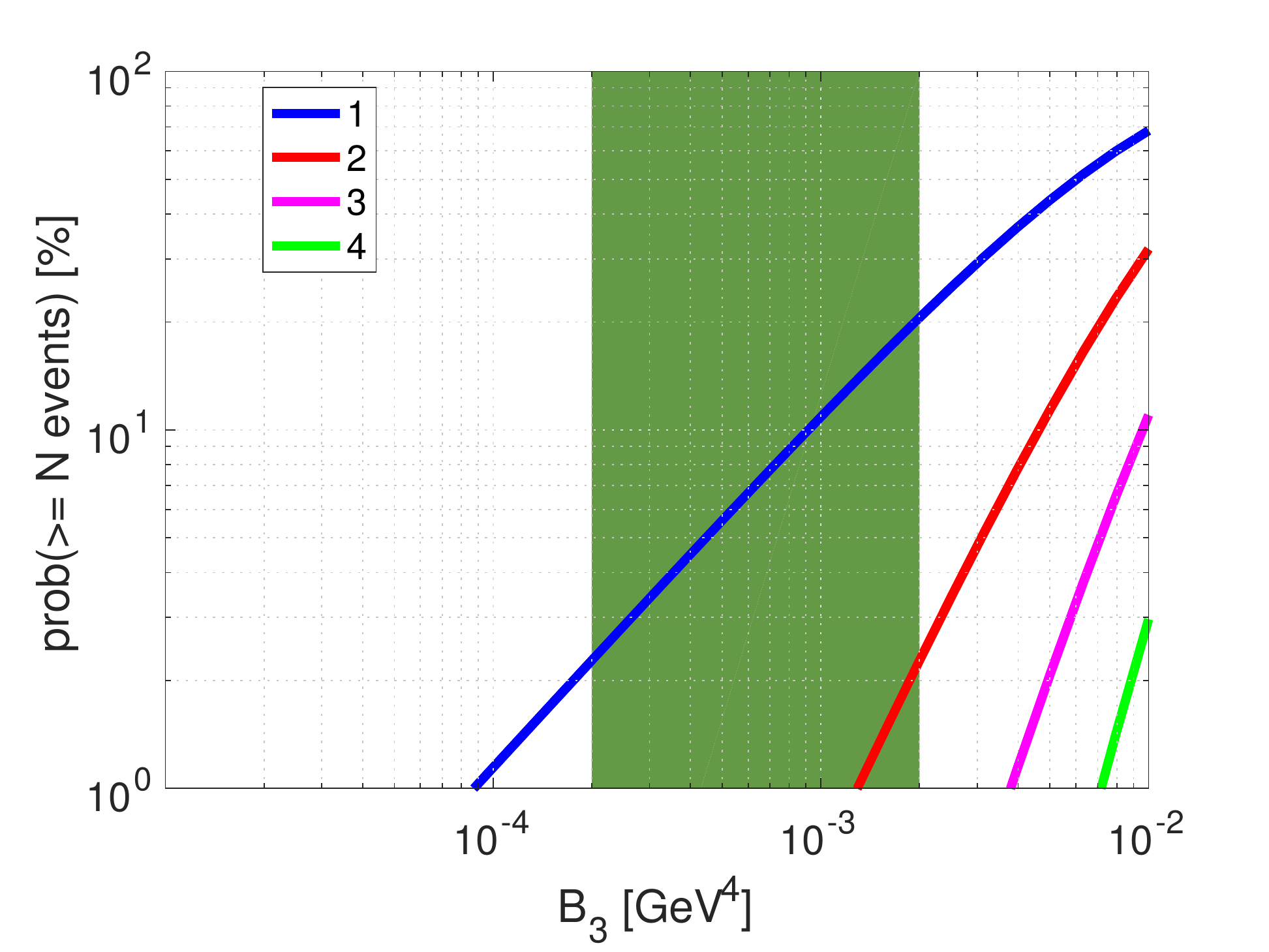}
\caption{Poisson probability for detecting $N\geq1,2,3,4$ $\ah$ events in a 5-yr analysis of AMS02, assuming the same exposure as in the $\bar p$ analysis~\cite{Aguilar:2016kjl}. Eq.~(\ref{eq:B3val}) shown as green band.}
\label{fig:Poiss}
\end{center}
\end{figure}

The resulting $\ah$ flux is plotted in Fig.~\ref{fig:flux}. A pre-launch estimate of the 18-yr 95\%CL $\ah$ flux upper limit accessible with AMS02 was given in Ref.~\cite{kounineHebar} in terms of the $\ah/$He flux ratio. In Fig.~\ref{fig:flux}, in dashed line, we plot this expected upper limit sensitivity, scaled to 5-yr exposure and multiplied by the observed He flux~\cite{Aguilar:2015ctt}. We learn that AMS02 may indeed detect secondary $\ah$ in a 5-yr analysis.
To further quantify this result, in Fig.~\ref{fig:Poiss} we show the Poisson probability for a 5-yr analysis of AMS02 to detect $N\geq1,2,3,4$ $\ah$ events. The calculation assumes an $\ah$ analysis with the same exposure as the 5-yr $\bar p$ analysis of~\cite{Aguilar:2016kjl}.\\ 

\noindent
\mysection{CR anti-deuterium}\label{sec:ppd}
\noindent
The $\bar d$ analysis is analogous to that of $\ah$. The $\bar d$ flux is plotted in Fig.~\ref{fig:flux}, for the range of $B_2$ given in Eq.~(\ref{eq:B2val}). 
AMS02 5-yr 95\%CL $\bar d$ flux sensitivity in the kinetic energy range 2.5-4.7~GeV/nuc, estimated in Ref.~\cite{Aramaki:2015pii}, is shown by solid lines. \\

\noindent
\mysection{Summary}\label{sec:conc}
\noindent
We calculate the flux of secondary cosmic ray $\bar p,\,\bar d$ and $\ah$. Propagation details are irrelevant to the calculation as long as consistent input data, notably B/C and proton flux, are used to calibrate it. The challenge is in deriving the correct production cross section in pp collisions, the dominant astrophysical source, for which accelerator data is scarce.

Using a scaling law of coalescence with HBT data we derive a novel estimate of the yield of $\ah$ in pp collisions. Our results are consistent with preliminary pp data from ALICE, motivating a dedicated analysis of $B^{(pp)}_3$ by the collaboration itself. Direct $\bar d$ data in pp collisions are also consistent with the HBT scaling.

Our prediction for the $pp\to\ah$ cross section is larger by 1-2 orders of magnitude compared to most previous estimates in the literature. The astrophysical secondary flux of $\ah$ is potentially within reach of a five-year exposure of AMS02.\\

\mysections{Acknowledgments}
We thank Ulrich W. Heinz and Yosef Nir for reading a draft version of this work, Urs Wiedemann for useful discussions and Andrei Kounine and Natasha Sharma for clarifying experimental details pertaining to AMS02 and ALICE analyses. This research is supported by the  I-CORE program of the Planning and Budgeting Committee and the Israel Science Foundation (grant number 1937/12). The work of MT is supported by the JSPS Research Fellowship for Young Scientists. KB is supported by grant 1507/16 from the Israel Science Foundation and is incumbent of the Dewey David Stone and Harry Levine career development chair. \\

\bibliography{ref}
\bibliographystyle{utphys}

\clearpage

\begin{appendix}
\noindent
\mysection{Appendix A: accelerator data for nuclear yield and HBT radius}
\noindent
The accelerator data analysis reported in Fig.~\ref{fig:Bformula} is summarised below. In all analyses, systematic uncertainties are important, related to the use of imperfect $pp\to\bar p$ (and, where required, $pp\to\pi$) cross section parametrisation, the treatment of threshold effects, and in some cases to the lack of complete information from the experiments. We do not attempt statistical fits but merely compare the data with Eq.~(\ref{eq:coal}) to extract estimates of $B_2$ and $B_3$.

We consider only data corresponding to $\bar d$, $\ah$, and $\at$ production. Low CME data such as~\cite{Saito:1994tg} (pA) and~\cite{Nagamiya:1981sd, Wang:1994rua,Barrette:1994tw, Armstrong:2000gd,Armstrong:2000gz,Bennett:1998be,Albergo:2002gi} (AA) demonstrates that production of $d$ and $^3$He necessarily involves additional processes that are different from the strongly inelastic formation of anti-nuclei that is of interest to us here. In particular, the approximate CME-independence of $B_2$ and $B_3$ that is found in high CME experiments is broken for nuclei production very near and below threshold (see e.g. Fig.~28 in Ref.~\cite{Anticic:2016ckv}).

\begin{figure}[t]
\begin{center}
\includegraphics[scale=0.35]{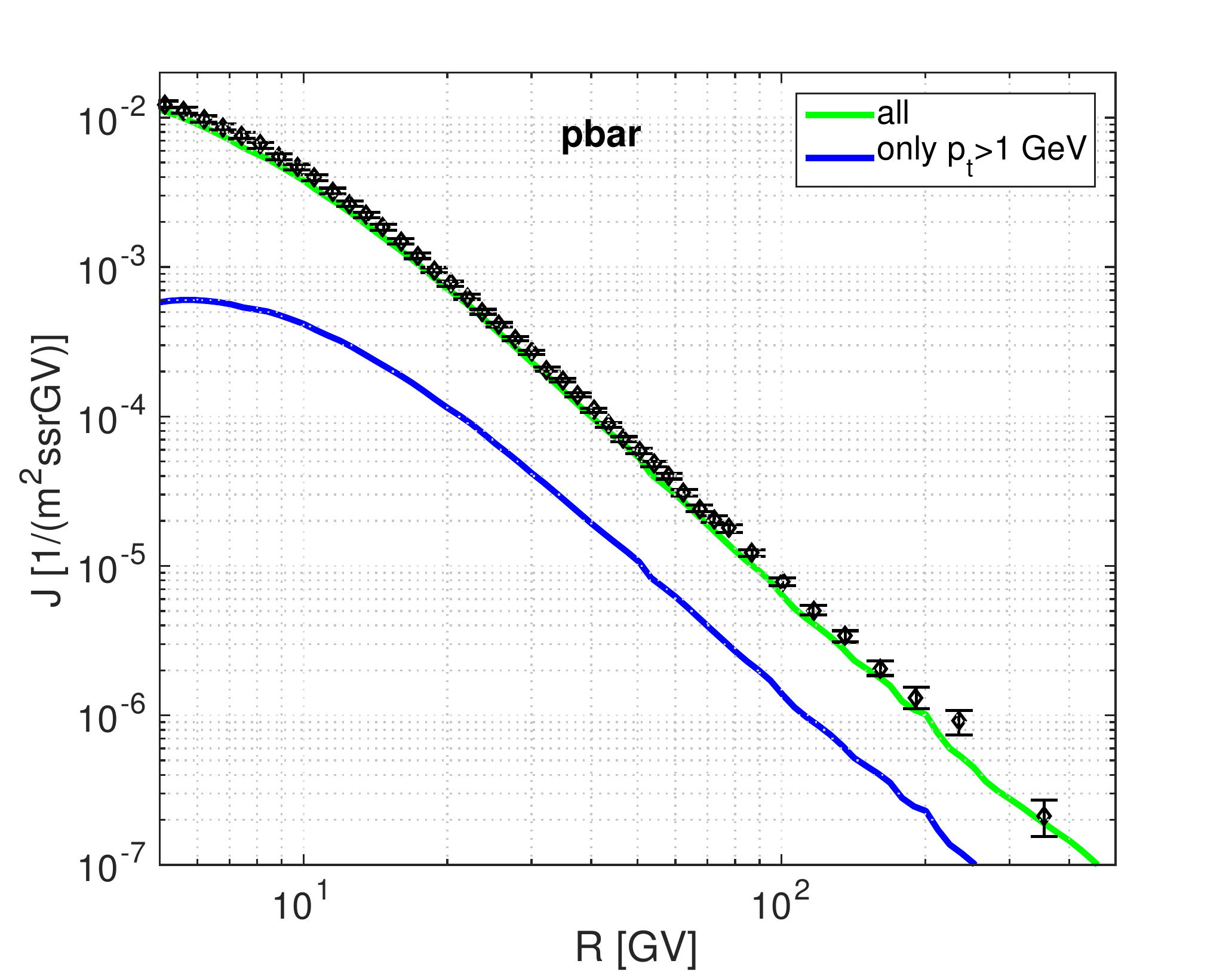}\quad
\includegraphics[scale=0.35]{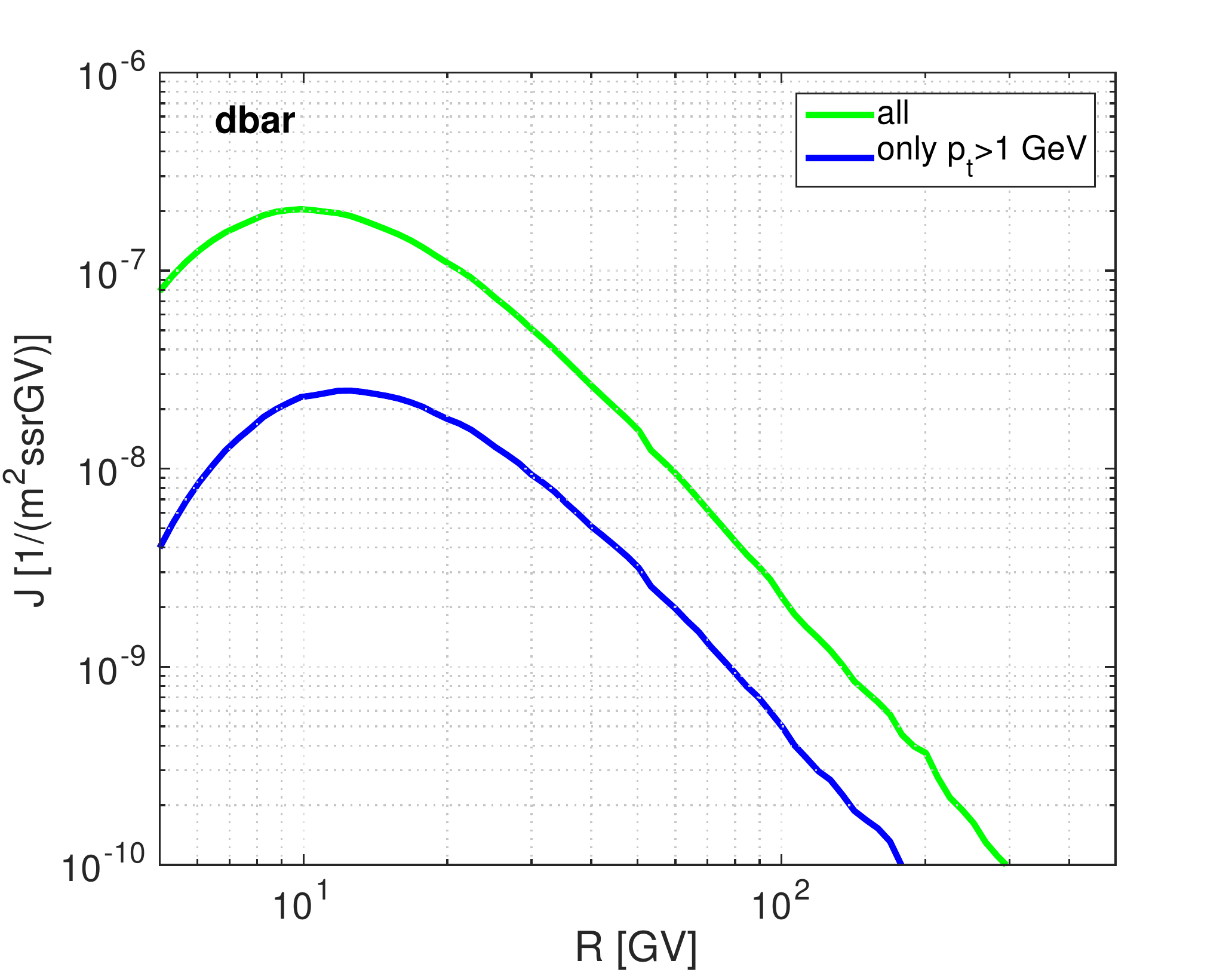}\quad
\includegraphics[scale=0.35]{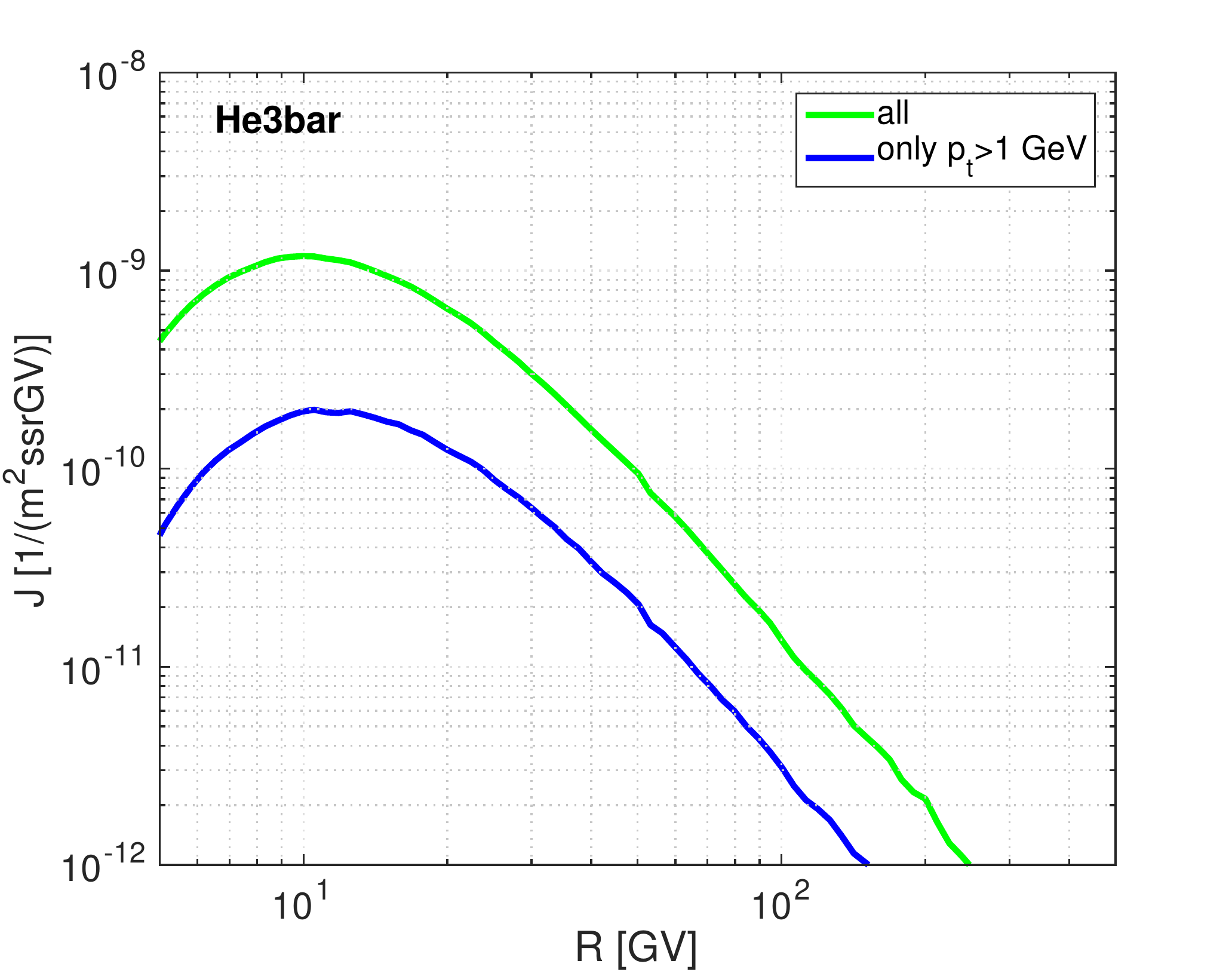}
\caption{Contribution of high-$p_t$ region to the secondary astrophysical production of CR anti-nuclei. $\bar d$ and $\ah$ curves use $B_2=1.4\times10^{-2}$~GeV$^2$ and $B_3=1.9\times10^{-3}$~GeV$^4$, respectively.}
\label{fig:ptCR}
\end{center}
\end{figure}

\begin{figure}[t]
\begin{center}
\includegraphics[scale=0.35]{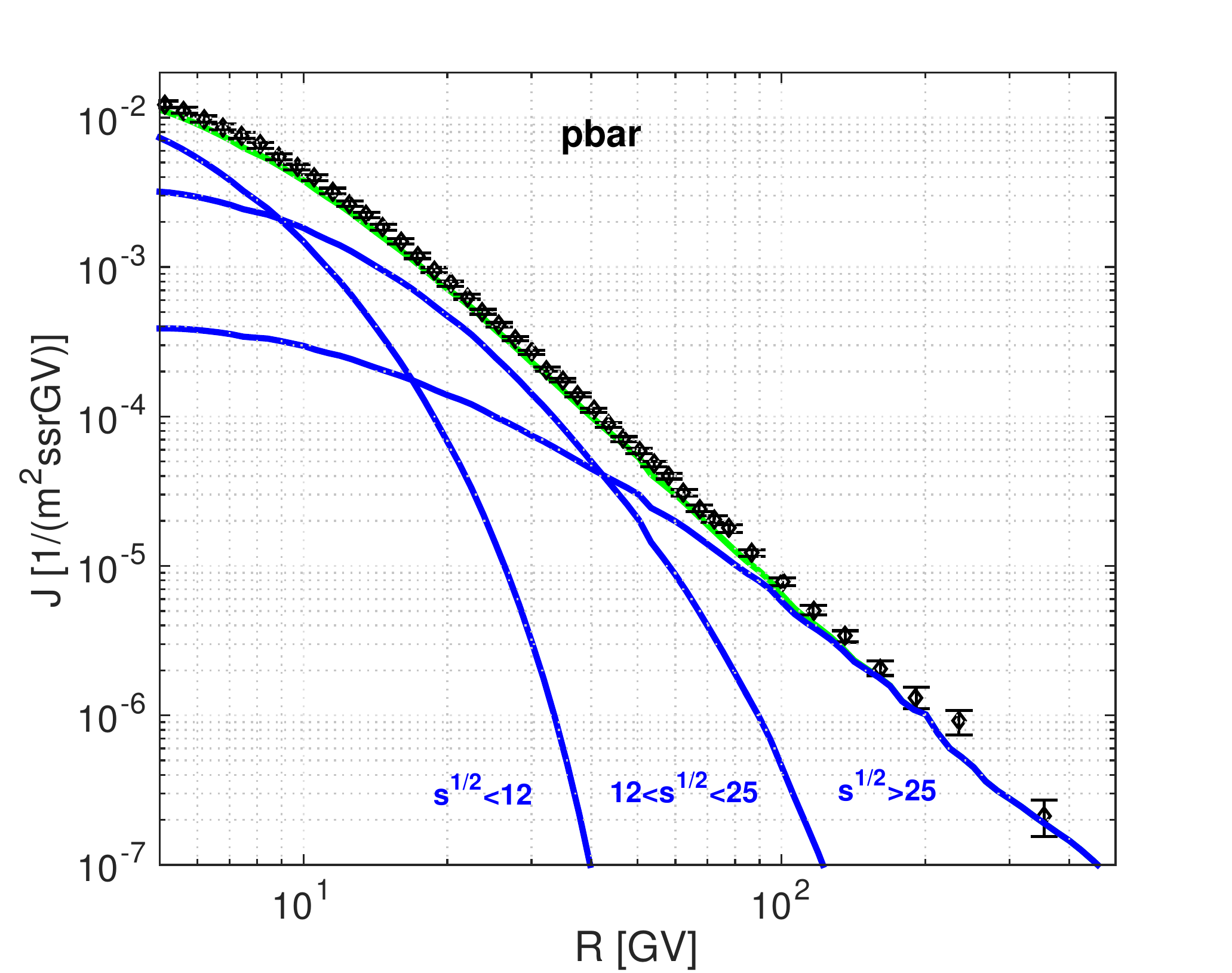}\quad
\includegraphics[scale=0.35]{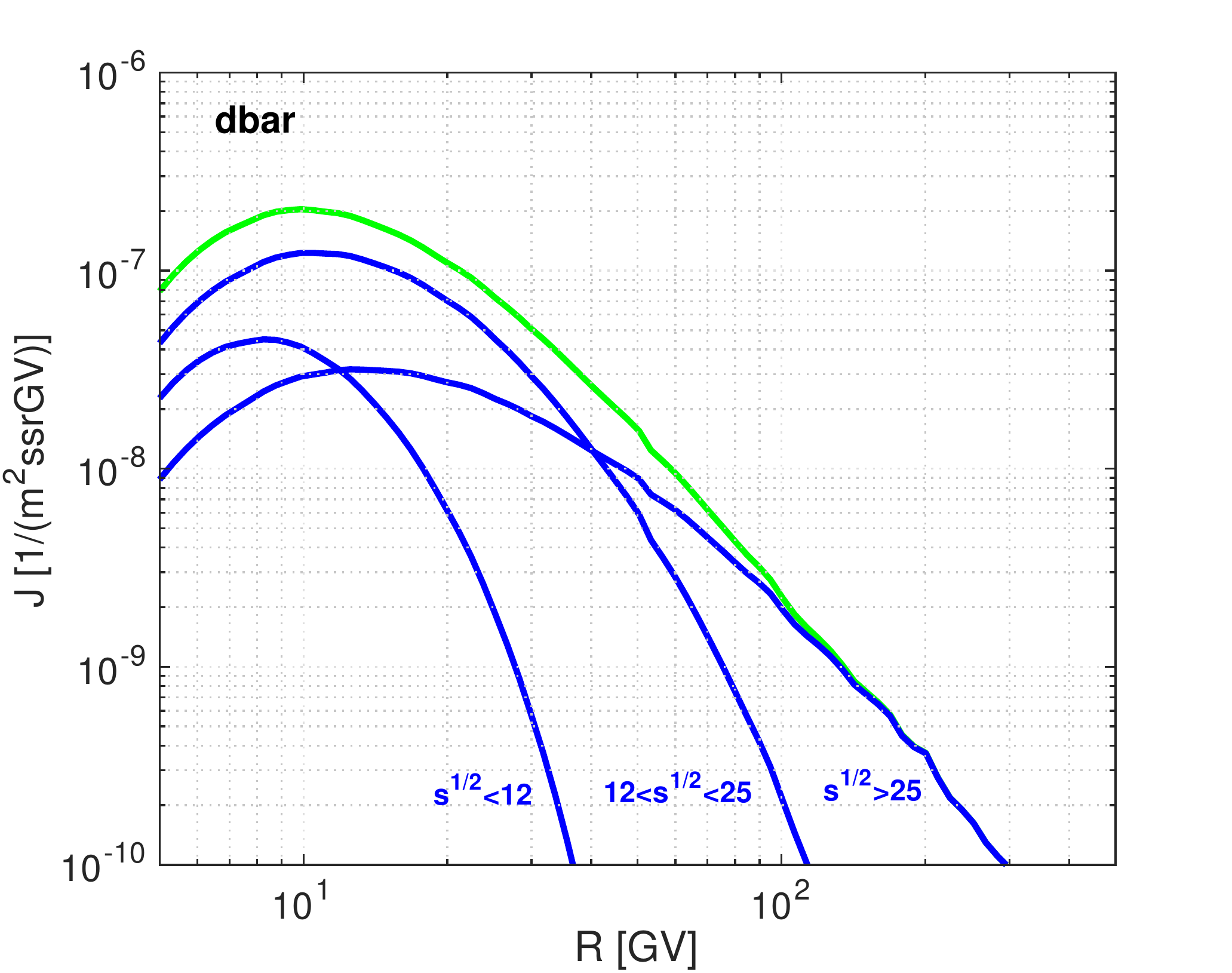}\quad
\includegraphics[scale=0.35]{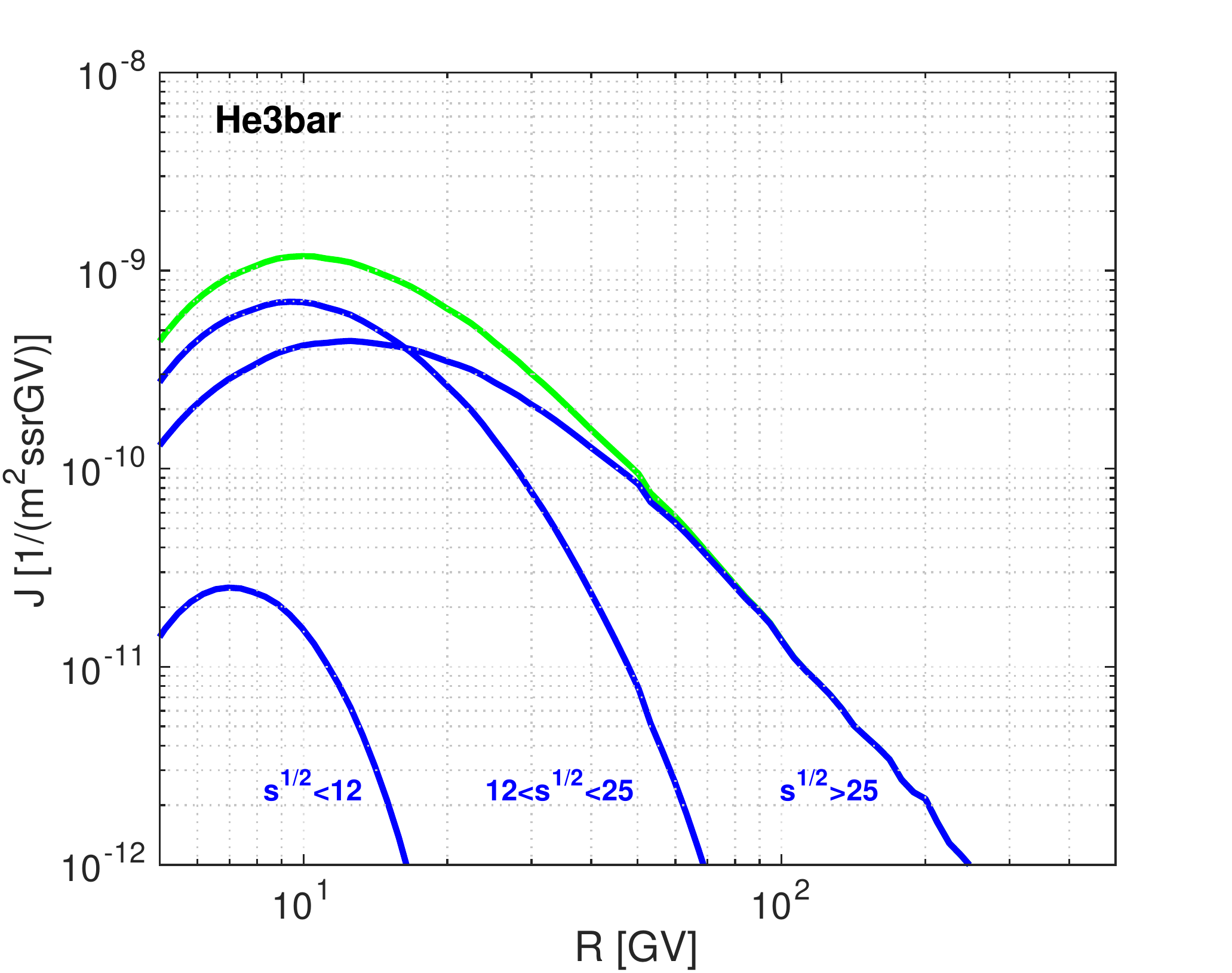}
\caption{Contribution of different $\sqrt{s}$ collisions to the secondary astrophysical production of CR anti-nuclei. $\bar d$ and $\ah$ curves use $B_2=1.4\times10^{-2}$~GeV$^2$ and $B_3=1.9\times10^{-3}$~GeV$^4$, respectively. $\sqrt{s}$ values in GeV.}
\label{fig:sCR}
\end{center}
\end{figure}

Before proceeding to details, we note that different data sets probed different kinematical regions, and it is useful to understand which kinematical regions dominate the astrophysical secondary source. A first rule of thumb is that only low-$p_t$ data is directly relevant to the astrophysics. We demonstrate this point by computing the CR flux, restricting the source term to anti-nucleus $p_t$ larger than 1~GeV. The result is shown in Fig.~\ref{fig:ptCR}. As can be seen, the anti-nucleus CR flux receives only a small contribution from the range $p_t>1$~GeV.
Second, in Fig.~\ref{fig:sCR} we plot the contribution to CR flux coming from pp collisions at different ranges of CME. 
\\

\underline{We turn to Fig.~\ref{fig:Bformula}. For the HBT radius $R$:}
\begin{itemize}
\item
{\bf pp Avg E766 30 GeV (R)}: Ref.~\cite{Uribe:1993tr} reported averaged pion HBT radius in fixed target pp collisions
  at proton momentum of $27.5$~GeV.
  Their analysis indicates $R\simeq 0.8-1.2$~fm.
\item
{\bf pp ALICE 7 TeV (R)}: Ref.~\cite{Malinina:2013fhp} reported kaon HBT radius in pp collisions at $\sqrt{s}=7$~TeV. The HBT radius weakly depends on transverse mass $m_t$ for $m_t\gtrsim1$~GeV.
We use the HBT radius data at low charged particle multiplicity and $m_t\simeq 1$~GeV since
such a radius is expected to represent the proton HBT radius at lower $\sqrt{s}$. We find $R\approx0.5-1.1$~fm. In Fig.~\ref{fig:Bformula} we join this result to that found above from~\cite{Uribe:1993tr}.
\item
{\bf pPb Avg NA44 450 GeV (R)}: Ref~\cite{Boggild:1998dx} reported the averaged proton HBT radius in pPb fixed-target 
experiment at proton momentum of 450~GeV. They give $R=1.25-1.58$~fm, which we use in Fig.~\ref{fig:Bformula} to describe pA systems.
\item
{\bf PbPb Central/Off ALICE 2.76 TeV with high/low $p_t$ (R)}: Ref.~\cite{Adam:2015vja} reported the anti proton HBT radius
in PbPb collision at $\sqrt{s_{\rm NN}}=2.76$ TeV.
The anti proton HBT radius weakly depends on $p_t$.
We pick up four types of the HBT radius:
  two centrality classes (central: $0-10\%$ and off: $30-50\%$)
and two transverse momenta (high: $p_t\simeq1.4$~GeV and low: $p_t\simeq0.6$~GeV).
\item 
{\bf PbPb Central NA44/NA49 158A GeV (R)}: Ref.~\cite{Bearden:2001sy} reported the kaon HBT radius
in central PbPb collision at $158$~GeV/nucleon. 
We use $m_t\simeq 1$~GeV data.
\item
{\bf AuAu Central STAR 200 GeV (R)}: Ref.~\cite{Chajecki:2005zw} reported the proton HBT radius in central AuAu collision
at $\sqrt{s_{\rm NN}}=200$~GeV. We use data at $p_t=0.6$ GeV.
\end{itemize}

\underline{For the derivation of $B_2$ (upper panel of Fig.~\ref{fig:Bformula}):}
\begin{itemize}
\item
{\bf pp ISR 53 GeV (B$_2(\bar{d})$)}: Ref.~\cite{ALBROW1975189,1973PhLB...46..265A,Henning:1977mt} reported $\bar d$ production in pp collisions at $\sqrt{s}=53$~GeV. Fig.~\ref{fig:antid_pp53GeV} summarises the analysis of~\cite{1973PhLB...46..265A,Henning:1977mt} data, which is the range that we quote in Eq.~(\ref{eq:B2val}). 

The high rapidity data from~\cite{ALBROW1975189} requires care, as our baseline $\bar p$ production cross section from Tan\&Ng~\cite{0305-4616-9-10-015} is inaccurate in this kinematical regime. Ref.~\cite{Albrow:1973qz} reported $pp\to\bar p$ cross sections in the same kinematical regime, where~\cite{0305-4616-9-10-015} over-estimates the data by a factor of 2-3. (That would lead to a factor of 4-9 underestimate in $B_2$.) The lowest $p_t$ $pp\to\bar p$ cross section measurement from~\cite{Albrow:1973qz} is at $\bar p$ $p_t$ of 0.15~GeV, allowing us to analyze the highest $p_t$ data point in~\cite{ALBROW1975189} (out of a grand total of 3 data points), where the $\bar d$ $p_t$ is 0.3~GeV. With this, we find $B_2^{pp}(\bar d)=(0.7-1.7)\times10^{-2}$~GeV$^2$, consistent with Eq.~(\ref{eq:B2val}). 

We do not analyze the ISR data from~\cite{ARMITAGE197987}, following the discussion in~\cite{Duperray:2002pj} that pointed out to potential inconsistency in the cross section reported for different values of $p_t$. A quick check, however, suggests that $B_2$ inferred from~\cite{ARMITAGE197987} is roughly consistent with the other ISR data we analyzed.
\begin{figure}[t]
\begin{center}
\includegraphics[scale=0.4]{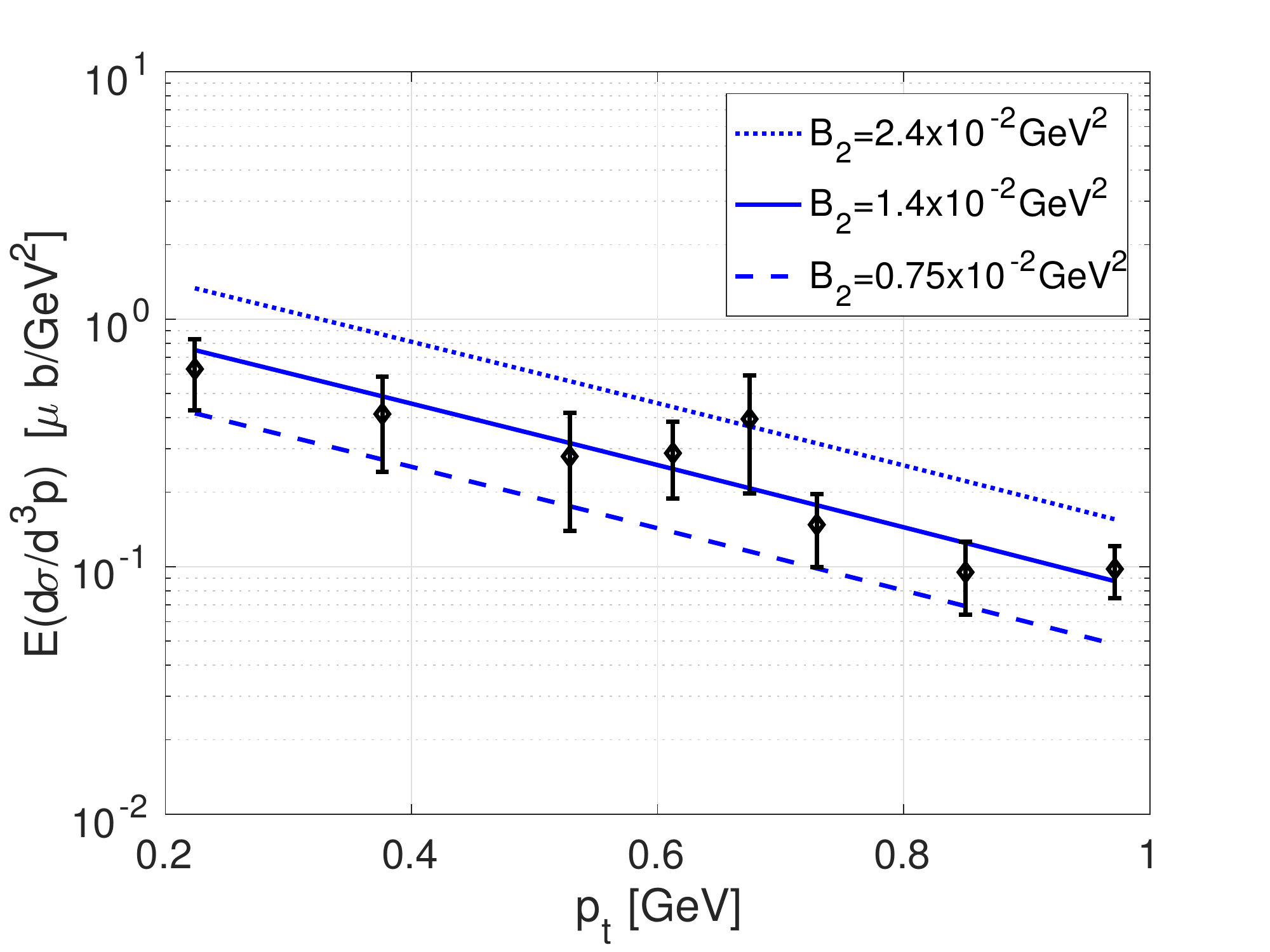}
\caption{$B^{(pp)}_2$ from ISR $\sqrt{s}=53$~GeV $pp\to\bar d$ data~\cite{1973PhLB...46..265A,Henning:1977mt}, using $\bar p$ and $\pi^-$ cross sections from~\cite{0305-4616-9-10-015} subtracting a 19\% hyperon contribution.}
\label{fig:antid_pp53GeV}
\end{center}
\end{figure}
\item 
{\bf pp Serpukhov 11.5 GeV (B$_2(\bar{d})$)}: Ref.~\cite{Abramov:1986ti} reported $\bar d$ production in pp collisions at $\sqrt{s}=11.5$~GeV. This data set is both low CME and high-$p_t$. The Tan\&Ng~\cite{0305-4616-9-10-015} $pp\to\bar p$ cross section fits are consistent (to better than 30\%) with the $pp\to\bar p$ data reported in~\cite{Abramov:1979hr} for the same set-up at $p_t<0.8$~GeV, but overestimate the data at higher $p_t$ with a factor of 7 discrepancy at $p_t=2.2$~GeV. (That would lead to a factor of 50 underestimate in $B_2$.) In Fig.~\ref{fig:Serpukhov} we analyze~\cite{Abramov:1986ti} using $pp\to\bar p$ data from~\cite{Abramov:1979hr}.  

Refs.~\cite{Duperray:2005si,Cirelli:2014qia} discarded~\cite{Abramov:1986ti} from their analyses. Indeed, our $B_2$ derived from~\cite{Abramov:1986ti} is significantly lower than that found from the ISR~\cite{ALBROW1975189,1973PhLB...46..265A,Henning:1977mt}. While we don't see any obvious reason to exclude~\cite{Abramov:1986ti}, we note that it corresponds to a kinematical regime that is not directly relevant to secondary CR $\bar d$ and $\ah$, as seen in Figs.~\ref{fig:ptCR}-\ref{fig:sCR}. 
For that reason we do not include the Serpukhov data in Eq.~(\ref{eq:B2val}).
\begin{figure}[t]
\begin{center}
\includegraphics[scale=0.4]{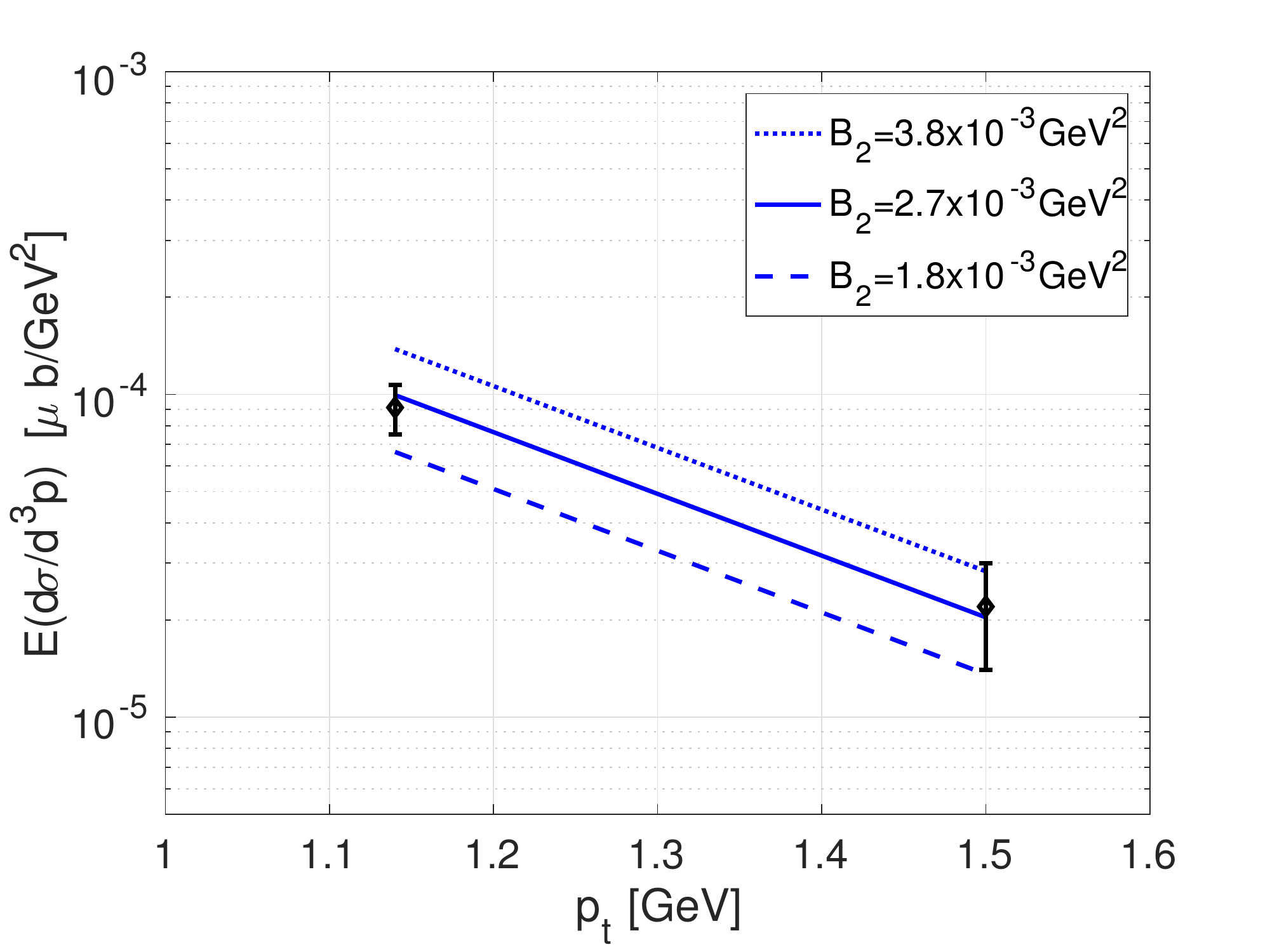}
\caption{$B^{(pp)}_2$ from Serpukhov $\sqrt{s}=11.5$~GeV $pp\to\bar d$ data~\cite{Abramov:1986ti}, analyzed using $pp\to\bar p$ data from~\cite{Abramov:1979hr} subtracting 19\% hyperon contributon.
}
\label{fig:Serpukhov}
\end{center}
\end{figure}
\item
{\bf pAl/Be SPS 200-240 GeV (B$_2(\bar{d})$)}: CERN SPS pBe and pAl fixed-target $\bar d$ data in the forward direction and at proton momentum of 200, 210, 240~GeV were reported in Ref.~\cite{BOZZOLI1978317,BUSSIERE19801}. The nucleon-nucleon CME energy is $\sqrt{s}\approx19.4,\,19.9,\,21.3$~GeV. (Note that production of $\bar d$ at rest in the centre of mass frame corresponds to lab frame $\bar d$ momentum $p_{lab}\approx20$~GeV. Lower/higher $p_{lab}$ means backward/forward momentum in the centre of mass frame). There are no absolute cross section measurements, but rather relative yields, e.g. $\bar d/\pi^-$ ratio in given momentum bins. We find that $\bar p/\pi^-$ data from~\cite{BUSSIERE19801} are reasonably well described (to $\sim30$\%) by the Tan\&Ng cross section fits~\cite{0305-4616-9-10-015}. To analyze the data we multiply the ratio $\bar d/\pi^-$ by the $pp\to\pi^-$ cross section and  divide by the relevant power of $pp\to\bar p$ evaluated at $p_{\bar p}=p_{\bar d}/2$ to extract $B_2$. Note that the need to use $pp\to\bar p,\pi^-$ as part of the process to extract $B_2$ (as opposed to deriving $B_2$ directly from the SPS pA set-up, for which, however, we are not given sufficient information) may lead to additional systematic uncertainty. The analysis is summarised in Fig.~\ref{fig:Bussiere1980}. 
\begin{figure}[t]
\begin{center}
\includegraphics[scale=0.4]{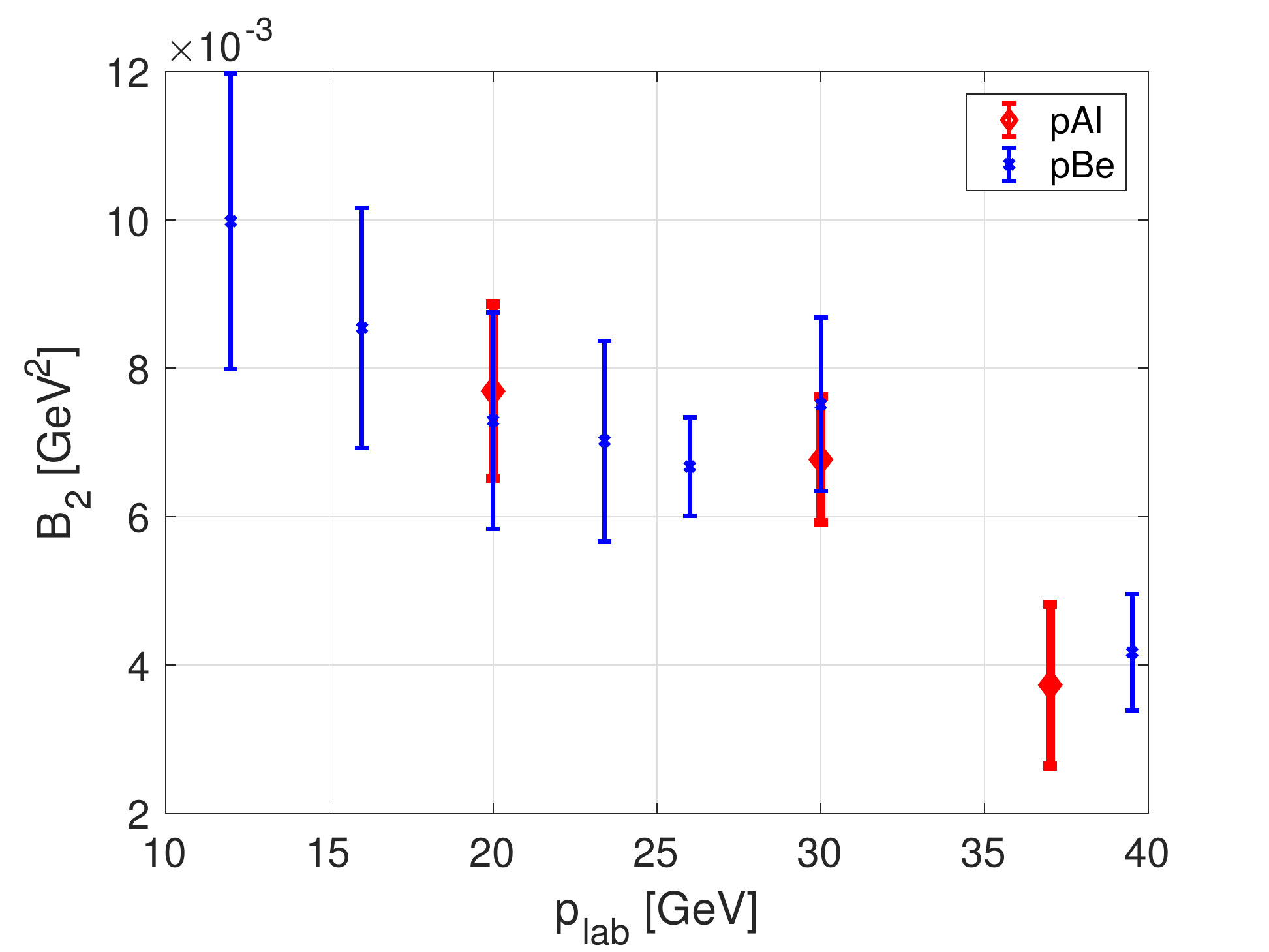}
\caption{$B^{(pA)}_2$ from SPS, $p_{lab}=200-240$~GeV, $\bar d$ data~\cite{BOZZOLI1978317,BUSSIERE19801}. We show the inferred value of $B_2$, with error bars reflecting only the quoted experimental uncertainties on the $\bar d/\pi^-$ ratio in the given hadron momentum bins. We use $pp\to\bar p,\pi^-$ cross sections from~\cite{0305-4616-9-10-015} subtracting a 19\% hyperon contribution.}
\label{fig:Bussiere1980}
\end{center}
\end{figure}
\item 
{\bf FNAL pBe/Ti/W 300~GeV (B$_2(\bar d)$)}: Ref.~\cite{Cronin:1974zm} reported large-angle $\bar d$ production in pBe/Ti/W collisions at 300~GeV incident p momentum. The kinematical regime of this data set, $p_t>2$~GeV, makes a negligible contribution to the astrophysical source, and $\bar p/\pi^-$ ratios from this measurement are not reproduced by~\cite{0305-4616-9-10-015}. Using a combination of $\bar d/\bar p,\,\bar p/\pi^-$ ratios and $\pi^-$ production cross sections from the same data set, we derive $B_2^{pBe}(\bar d)=(1.5-2.4)\times10^{-2}$~GeV$^2$, $B_2^{pTi}(\bar d)=(2.8-4.1)\times10^{-2}$~GeV$^2$, $B_2^{pW}(\bar d)=(3.2-3.8)\times10^{-2}$~GeV$^2$, for $p_t=2.29$~GeV. These results are then multiplied by a factor of 1.5 to account for hyperon contribution to the $\bar p$ cross section in those analyses.
\item
{\bf PbPb Central/Off ALICE 2.76 TeV with high/low $p_t$ (B$_2(\bar{d})$, B$_2(d)$)}: Ref.~\cite{Adam:2015vda}
reported $B_2$ for $d$ and $\bar d$ in PbPb collision at $\sqrt{s_{\rm NN}}=2.76$ TeV. The data shows $B_2(\bar d)\approx B_2(d)$, with weak dependence on $p_t$.
We pick up four types of $B_2$:
 two centrality classes (central: $0-10\%$ and off: $20-60\%$)
 and two transverse momenta (high: $p_t/2\simeq1.4$~GeV and low: $p_t/2\simeq0.6$~GeV).
\item
{\bf PbPb Central NA44/NA49 158A GeV (B$_2(d)$)}: Ref.~\cite{Anticic:2016ckv} reported
$B_2$ for $d$
in central PbPb collision at $158$~GeV/nucleon.
$B_2$ weakly depends on $p_t$ and we take the $p_t\simeq 0$ data point.
\item
{\bf AuAu Central STAR 200 GeV (B$_2(\bar{d})$, B$_2(d)$)}: Ref.~\cite{Liu:2008zzi} reported $B_2$ for d and $\bar d$
in central AuAu collision
at $\sqrt{s_{\rm NN}}=200$~GeV.
The data shows $B_2(\bar d)\approx B_2(d)$, with weak dependence on $p_t$. 
We take the $p_t/2\simeq 0.8$ GeV data point.
\end{itemize}

\underline{For the derivation of $B_3$ (lower panel of Fig.~\ref{fig:Bformula}):}
\begin{itemize}
\item
{\bf pAl/Be SPS 200-240 GeV (B$_3(\ah)$, B$_3(\at)$)}: The analysis of the $\ah$ and $\at$ data from Ref.~\cite{BOZZOLI1978317,BUSSIERE19801} is analogous to that described for $\bar d$. The systematic uncertainties here are more severe, because the $\bar p$ distributions are sampled at lower momentum and because they must be raised to a higher power to extract $B_3$. 
Our analysis is summarised in Fig.~\ref{fig:Bussiere1980He3bar}. In quoting the result in Fig.~\ref{fig:Bformula} we discard the highest $p_{lab}$ data point.
\begin{figure}[t]
\begin{center}
\includegraphics[scale=0.4]{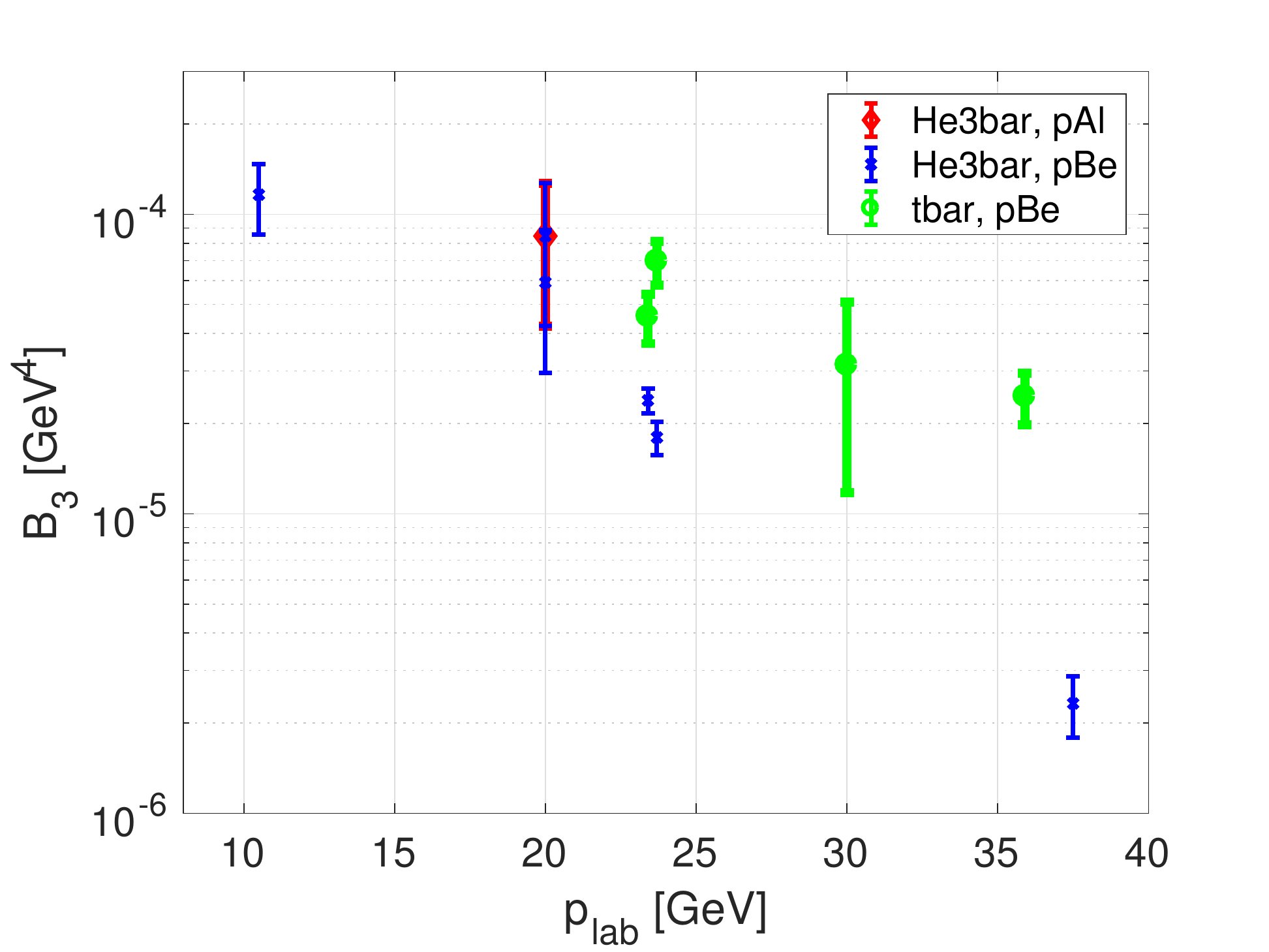}
\caption{$B^{(pA)}_3$ from SPS, $p_{lab}=200-240$~GeV, $\ah,\at$ data~\cite{BOZZOLI1978317,BUSSIERE19801}. We show the inferred value of $B_3$, with error bars reflecting only the quoted experimental uncertainties on the $\ah/\pi^-$ and $\at/\pi^-$ ratios in the given momentum bins. We use $pp\to\bar p,\pi^-$ cross sections from~\cite{0305-4616-9-10-015} subtracting a 19\% hyperon contribution.}
\label{fig:Bussiere1980He3bar}
\end{center}
\end{figure}
\item
{\bf PbPb Central/Off ALICE 2.76 TeV with high/low $p_t$ (B$_3(\ah)$, B$_3(^3\rm He)$)}: Ref.~\cite{Adam:2015vda} also
reported $B_3$ for both $^3\rm He$ and $\ah$ in PbPb collision at $\sqrt{s_{\rm NN}}=2.76$ TeV.
The data shows $B_3(\ah)\approx B_3(^3\rm He)$ with weak dependence on $p_t$.
We pick up four types of $B_3$:
 two centrality classes (central: $0-10\%$ and off: $20-80\%$)
 and two transverse momenta (high: $p_t/3\simeq1.4$~GeV and low: $p_t/3\simeq0.8$~GeV).
\item
{\bf PbPb Central NA44/NA49 158A GeV  (B$_3(^3\rm He)$)}: 
Ref.~\cite{Anticic:2016ckv} reported
$B_3$ for $^3\rm He$
in central PbPb collision at $158$~GeV/nucleon.
$B_3$ weakly depends on $p_t$ and we take the $p_t\simeq 0$ data point.

Ref.~\cite{0954-3899-23-12-039} reported $B_3^{PbPb}(\ah)=1^{+2}_{-\infty}\times10^{-6}$, $B_3^{PbPb}(^3{\rm He})=1^{+2}_{-\infty}\times10^{-5}$, $B_2^{PbPb}(\bar d)=(1-1.5)\times10^{-3}$, $B_2^{PbPb}(d)=(0.7-1.1)\times10^{-3}$. In addition to the PbPb measurements, AA and pA results from other experiments were also summarised, reporting $B_3^{pA}\approx10^{-4}$, $B_{2}^{pA}\approx10^{-2}$, similar for matter and antimatter. In particular, the SPS pAl/Be data of~\cite{BUSSIERE19801} is quoted as $B_3^{pA}(\ah)\approx B_3^{pA}(^3{\rm He})\approx2\times10^{-4}$. However, no derivation is reported, and the latter result -- while it agrees with expectations from Eq.~(\ref{eq:BR}) and with other data in Fig.~\ref{fig:Bformula} -- does not quite agree with what we find in our own analysis of~\cite{BUSSIERE19801}.
\item
{\bf AuAu Central STAR 200 GeV (B$_3(\ah)$, B$_3(^3\rm He)$)}:
Ref.~\cite{Liu:2008zzi} also reported $B_3$ for $^3\rm He$ and $\ah$
in central AuAu collision
at $\sqrt{s_{\rm NN}}=200$~GeV.
The data shows $B_3(\ah)\approx B_3(^3\rm He)$ with weak dependence on $p_t$.
We take the $p_t/3\simeq 0.8$ GeV data point.
\end{itemize}

Finally, we have also analyzed a number of intermediate CME pp and pA experimental results for matter. While this data appears to be generally consistent with the trend for antimatter, we do not add it to Fig.~\ref{fig:Bformula}, as it may lead to upward bias in $B_2$ and $B_3$. Very low CME nuclei production data~\cite{Saito:1994tg,Nagamiya:1981sd, Wang:1994rua,Barrette:1994tw, Armstrong:2000gd,Armstrong:2000gz,Bennett:1998be,Albergo:2002gi} shows that matter coalescence (or fragmentation, in this case) contains additional channels beyond those available for antimatter, that may contaminate the intermediate CME regime. For completeness we summarise our results below.
\begin{itemize}
\item 
{\bf pp Serpukhov 11.5 GeV (B$_2(d)$)}: Ref.~\cite{Abramov:1986ti} also reported the $d$ yield. We analyze the data using $pp\to p$ cross sections from~\cite{Abramov:1979hr}.  
We find $B_2(d)\sim B_2(\bar d)\times2.2$. As in the $\bar d$ data from the same reference, the data corresponds to low CME/ high-$p_t$.
\item
{\bf pAl/Be SPS 200-240 GeV (B$_2(d)$)}: Ref.~\cite{BOZZOLI1978317,BUSSIERE19801} also reported the yield of deuterons. The cross section fits of~\cite{0305-4616-9-10-015} are inaccurate for $pp\to p$, meaning that we cannot repeat our exercise for the $\bar d$ analysis of the same reference. Instead, we use Fig.~4 of~\cite{BUSSIERE19801} to estimate of $B_2(d)$, finding $B_2(d)\sim B_2(\bar d)\times2.5$. Note that Fig.~4 of~\cite{BUSSIERE19801} uses a Hagedorn-Ranft model to translate relative hadron yields to cross section. However, the same model fails to reproduce the relative hadron yields in~\cite{BUSSIERE19801} (see Fig.~2 there). The estimate of $B_2(d)$ derived this way should therefore be taken with caution.
\item 
{\bf FNAL pBe/Ti/W 300~GeV (B$_2(d)$)}: Ref.~\cite{Cronin:1974zm} also reported $d$ production. Following a similar prescription as we did for the $\bar d$ data, we derive $B_2^{pBe}(d)=(0.9-1.3)\times10^{-2}$~GeV$^2$, $B_2^{pTi}(d)=(2-3)\times10^{-2}$~GeV$^2$, $B_2^{pW}(d)=(3.0-3.8)\times10^{-2}$~GeV$^2$, for $p_t=2.29$~GeV.
\item
{\bf pC/Al/Cu/W 50 GeV IHEP-SPIN (B$_2(d)$)}: Ref.~\cite{Antonov:2017uyi}
reported $B_2$ for $d$ in a collision of 50 GeV proton beam with targets of C, Al, Cu, and W:
$B_2=(1.1 - 1.5)\times 10^{-2}$ GeV$^2$ for $p_t \simeq 1.4$ GeV.
\item
{\bf pC/Al/Cu/W 50 GeV IHEP-SPIN (B$_3(t)$)}: Ref.~\cite{Antonov:2017uyi}
reported $B_3$ for $t$ in a collision of 50 GeV proton beam with targets of C, Al, Cu, and W, giving $B_3=(0.8 - 1.5)\times 10^{-4}$ GeV$^4$ for $p_t \simeq 1.8$ GeV.
\end{itemize}
%

\begin{figure}[t]
\begin{center}
\includegraphics[scale=0.42]{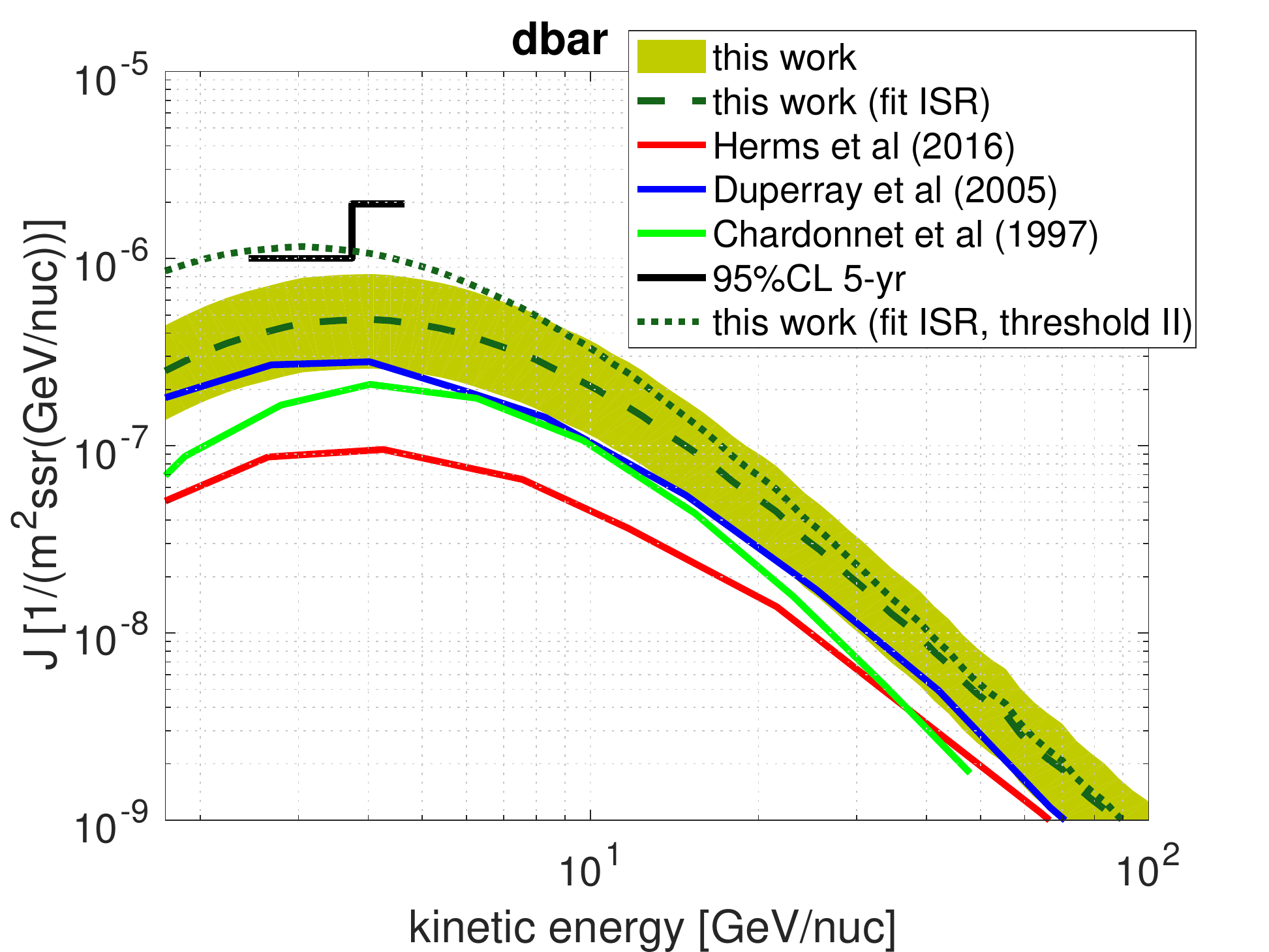}\quad
\includegraphics[scale=0.42]{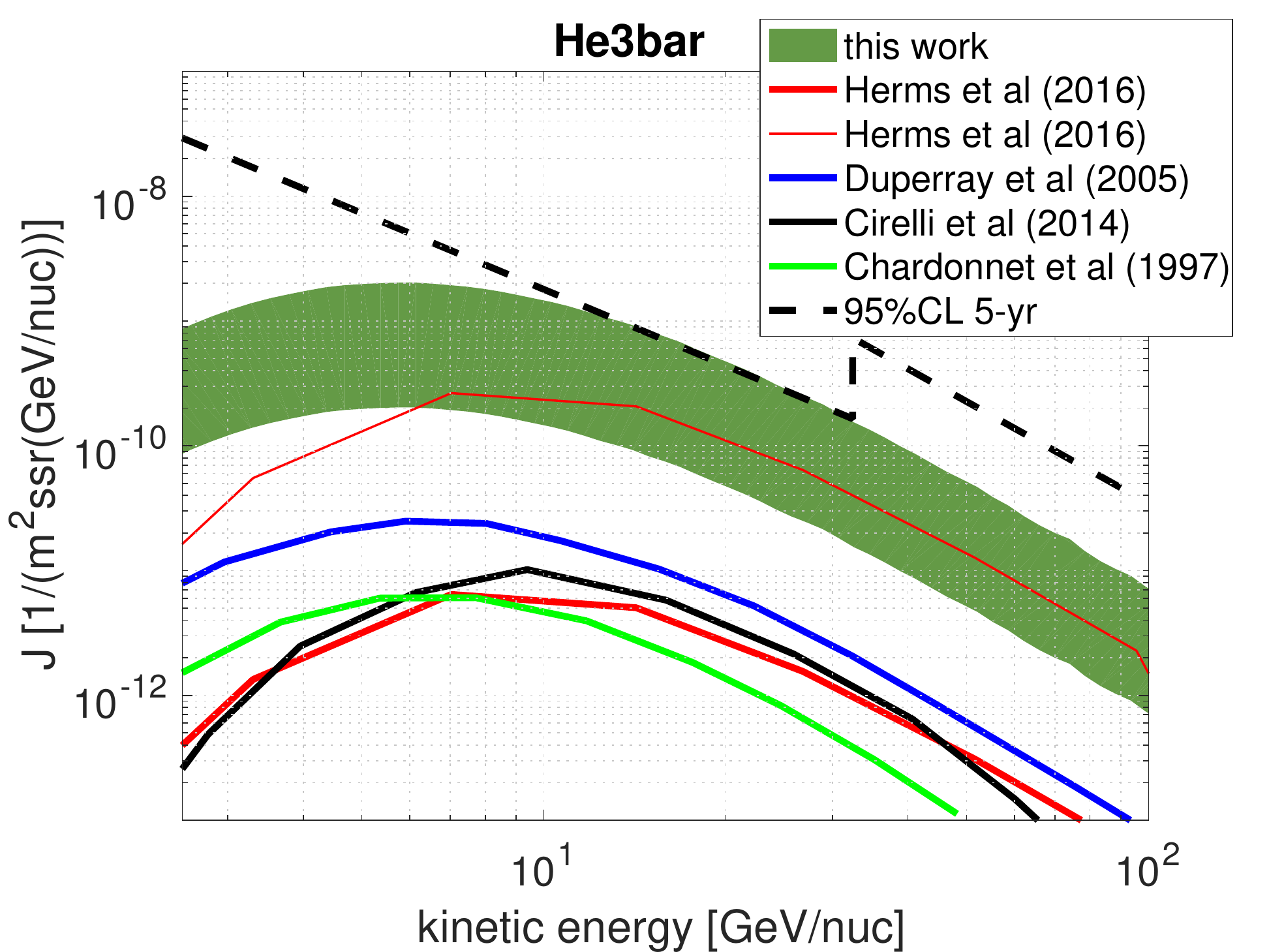}
\caption{Comparison to previous work. Top: $\bar d$ flux. Bottom: $\ah$ flux. We convert $\bar d/p$ and $\ah/p$ ratios from Chardonnet et al~\cite{Chardonnet:1997dv} using AMS02 p flux~\cite{Aguilar:2015ooa}. For Duperray et al~\cite{Duperray:2005si} we take the result including the direct $pp\to\ah$ channel. For Herms et al~\cite{Herms:2016vop}, the $\bar d$ prediction is the same as in~\cite{Ibarra:2013qt}; for $\ah$ the upper thin line corresponds to their estimate allowing $p_c(\ah)>p_c(\bar d)$.}
\label{fig:prev}
\end{center}
\end{figure}

\noindent
\mysection{Appendix B: Comparison to previous work}
\noindent
Our $\bar d$ and $\ah$ CR flux prediction is compared to previous work in the top and bottom panels of Fig.~\ref{fig:prev}, respectively. In what follows we present a detailed discussion of this comparison.
\underline{Chardonnet et al~\cite{Chardonnet:1997dv}} found $p_c=0.058$~GeV for $\bar d$, corresponding to $B_2\approx1.7\times10^{-3}$~GeV$^2$. The analysis included Serpukhov~\cite{Abramov:1986ti} and ISR~\cite{ALBROW1975189,1973PhLB...46..265A,Henning:1977mt} $pp\to\bar d$ data, and was based on $\bar p$ cross sections from~\cite{0305-4616-9-10-015}. The choice of $p_c$ was made to match the Serpukhov data, which yielded the lowest value of $p_c$ (and thus of $B_2$). We find that fitting $B_2$ to~\cite{Abramov:1986ti} gives a result that is lower by a factor of $\sim5$ than that needed to fit~\cite{1973PhLB...46..265A,Henning:1977mt}
\footnote{Our result for $B_2$ derived from the high rapidity ISR data~\cite{ALBROW1975189} is higher than that of~\cite{Chardonnet:1997dv} by a factor of about 7. This is explained by our use of different $pp\to\bar p$ cross section fits to analyze this data set. As explained in Appendix~A, the Tan\&Ng~\cite{0305-4616-9-10-015} fit over-estimates the high rapidity ISR $\bar p$ yield~\cite{Albrow:1973qz} by a factor of $\sim2-3$.}. 
As discussed in App.~A, we (and similarly~\cite{Duperray:2005si,Cirelli:2014qia}) use the ISR data~\cite{ALBROW1975189,1973PhLB...46..265A,Henning:1977mt} in Eq.~(\ref{eq:B2val}), rather than the value found based on the low CME, high-$p_t$ Serpukhov data~\cite{Abramov:1986ti}. To fully compare our $\bar d$ flux with~\cite{Chardonnet:1997dv}, we need to modify our phase space factor $R(x)$ to match their different prescription. We show the result in dotted line in Fig.~\ref{fig:prev}, reproducing the expected factor of 5 between our $\bar d$ flux and that of~\cite{Chardonnet:1997dv}.

For $\ah$, Ref.~\cite{Chardonnet:1997dv} simply used the same value of $p_c$ obtained in the $\bar d$ analysis to derive $B_3\approx2.3\times10^{-6}$~GeV$^4$. In addition, direct $\ah$ production was neglected and only the $pp\to\at$ channel was considered. The net result is a CR $\ah$ flux lower by a factor of $\sim100$ compared to our prediction.

\underline{Duperray et al~\cite{Duperray:2005si}} (following~\cite{Duperray:2002pj}) found $p_c=0.079$~GeV for $\bar d$, corresponding to $B_2\approx4.4\times10^{-3}$~GeV$^2$. The analysis collected together pp and pA data sets in a single statistical fit of $p_c$. As Fig.~\ref{fig:Bformula} and Eq.~(\ref{eq:BR2}) suggest, this could pull the fit artificially to low $B_2$, if the underlying physics satisfies $B_2^{pA}<B_2^{pp}$. 
In fact, considering the ISR $pp\to\bar d$ data~\cite{1973PhLB...46..265A,Henning:1977mt}, the global fit of~\cite{Duperray:2005si} is systematically below the data by a factor of about 2 (see Fig.~1 in~\cite{Duperray:2005si}). Restoring the factor of 2 gives a result consistent with our Eq.~(\ref{eq:B2val}) and, given the modest difference in $pp\to\bar p$ parametrisation, reproduces the difference between our $\bar d$ flux and that of~\cite{Duperray:2005si}.

Some more details: for the high rapidity ISR data~\cite{ALBROW1975189}, we reproduce the result of~\cite{Duperray:2005si} using their $pp\to\bar p$ cross section parametrisation, but we find that that parametrisation overestimates the $pp\to\bar p$ data of~\cite{Albrow:1973qz} by a factor of $\sim$2. This may explain why the fit of~\cite{Duperray:2005si} underestimates~\cite{1973PhLB...46..265A,Henning:1977mt} while at the same time slightly overestimating the highest $p_t$ data point of~\cite{ALBROW1975189}. The need for a careful treatment of $pp\to\bar p$ cross section in analyzing~\cite{ALBROW1975189} was also noted in~\cite{Duperray:2002pj}, who, however, extrapolated the $pp\to\bar p$ cross section fit derived in~\cite{Albrow:1973qz} to $p_t$ significantly lower than it was made to describe. Moving to pA data, our result for $B_2$ derived from~\cite{BOZZOLI1978317,BUSSIERE19801} agrees with~\cite{Duperray:2005si}. The main difference between our $\bar d$ analyses, therefore, is that we do not enforce $B_2^{pA}=B_2^{pp}$, such that the low $B_2$ derived from~\cite{BOZZOLI1978317,BUSSIERE19801} does not control our $B_2^{pp}$ result. 

For $\ah$, the $p_c=0.079$~GeV of~\cite{Duperray:2005si} translates into $B_3\approx1.5\times10^{-5}$~GeV$^4$, which was compared to the sparse data from~\cite{BOZZOLI1978317,BUSSIERE19801}. Our parallel analysis summarised in Fig.~\ref{fig:Bussiere1980He3bar} gives consistent results. However, as can be made clear by inspection of either of Fig.~\ref{fig:Bussiere1980He3bar} here, Fig.~4 in~\cite{Duperray:2005si}, or Tab.~2 in~\cite{BUSSIERE19801}, the sparse data leaves room for roughly an order of magnitude of systematic uncertainty in $B_3^{pA}$. 
The main difference between our $\ah$ analysis and that in~\cite{Duperray:2005si} is, therefore, the conclusion: Ref.~\cite{Duperray:2005si} made their prediction for $B_3^{pp}$ based on the (poor!) $B_3^{pA}$ fit to~\cite{BOZZOLI1978317,BUSSIERE19801}, while we (i) allow room for a factor of few uncertainty in the $\ah$ yield from~\cite{BOZZOLI1978317,BUSSIERE19801}, and (ii) expect $B_3^{pp}>B_3^{pA}$, based on the HBT scaling argument. Finally, another factor of $\sim2$ enhancing our predicted astrophysical $\ah$ flux compared to that of~\cite{Duperray:2005si} comes from a difference between our phase space factor $R_5(x)$ and that found by~\cite{Duperray:2005si}, at the relevant range $x\sim15-20$~GeV. The net outcome is a factor of $\sim50$ between our $\ah$ flux prediction and that of~\cite{Duperray:2005si}.

\begin{figure}[t]
\begin{center}
\includegraphics[scale=0.65]{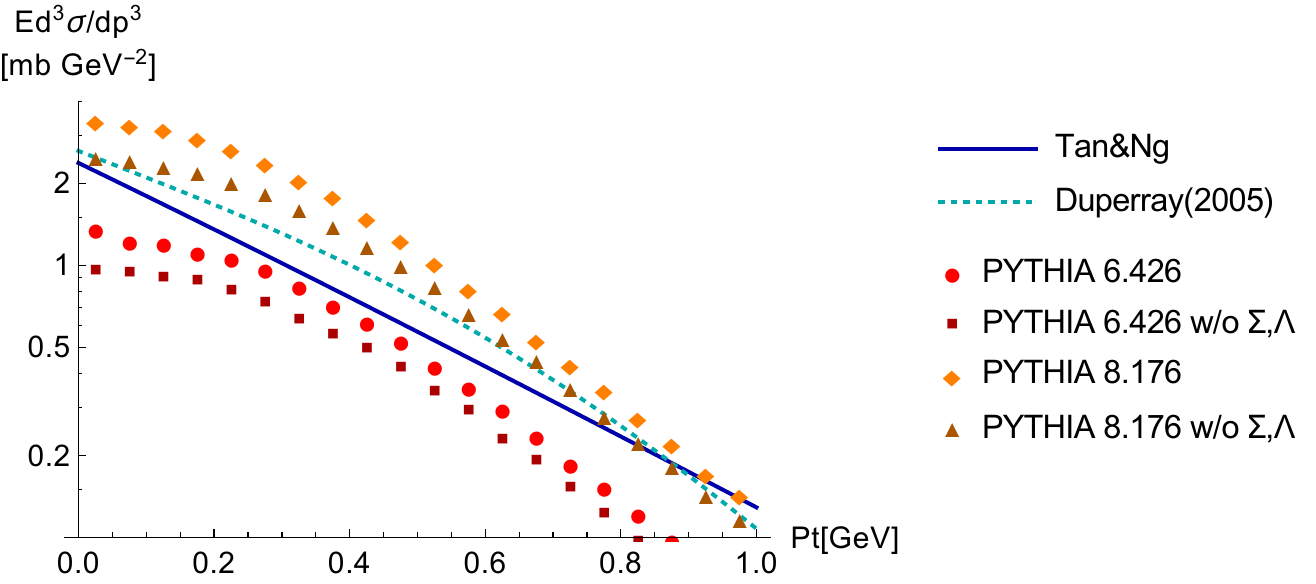}
\caption{Comparison of $pp\to\bar p$ production cross section parametrisation in the ISR set-up~\cite{1973PhLB...46..265A,Henning:1977mt}. Ref.~\cite{Cirelli:2014qia} chose to use PYTHIA 6.4.26 to calibrate their $\ah$ production.}
\label{fig:pythia}
\end{center}
\end{figure}

\underline{Cirelli et al~\cite{Cirelli:2014qia}} used PYTHIA 6.4.26~\cite{Sjostrand:2006za} to simulate $\bar p$ and $\bar n$ production in pp collisions;  fitted a coalescence momentum such that the calculation matches the ISR $pp\to\bar d$ data from Ref.~\cite{1973PhLB...46..265A,Henning:1977mt}; and then used that coalescence momentum to calculate $\ah$ production. The direct channel $pp\to\ah$ was dropped, and only the $pp\to\at$ channel  included. The basic building block, $pp\to\bar p,\,\bar n$ cross section, was not calibrated in~\cite{Cirelli:2014qia} to accelerator data, but simply taken from PYTHIA. 
We have made a comparison of the $pp\to\bar p$ cross section computed in PYTHIA 6.4.26 and by the Tan\&Ng~\cite{0305-4616-9-10-015} parametrisation, for the ISR set-up~\cite{1973PhLB...46..265A,Henning:1977mt}. The result is shown in Fig.~\ref{fig:pythia}. Note that Ref.~\cite{Cirelli:2014qia} subtracted the contribution of long-lived $\overline\Sigma$ and $\overline\Lambda$ baryons from the $\bar p$ cross section entering the definition of the coalescence fit. The codes differ by a factor of $\sim2$ in the relevant low $p_t$ region, with PYTHIA 6.4.26 lying below the Tan\&Ng~\cite{0305-4616-9-10-015} result which, however, does a fairly good job describing low $p_t$ ISR data. 
The end result is that the $\ah$ flux by~\cite{Cirelli:2014qia} is a factor of $\sim2$ below that of~\cite{Duperray:2005si}, and a factor of $\sim100$ below ours.

\underline{Ibarra \& Wild~\cite{Ibarra:2012cc,Ibarra:2013qt}} did the following exercise. First, in~\cite{Ibarra:2012cc} they used $pp\to\bar p$ cross sections from PYTHIA 8~\cite{Sjostrand:2007gs} to fit a coalescence model to ISR $pp\to\bar d$ data~\cite{1973PhLB...46..265A,Henning:1977mt}. 
Then, in~\cite{Ibarra:2013qt}, the same $p_c$ was plugged into DPMJET-III~\cite{Roesler:2000he} to calculate the CR $\bar d$ production. This time, an empirical $\sqrt{s}$-dependent correction factor was used to bring the original DPMJET-III $pp\to\bar p$ cross section into agreement with experimental data from~\cite{Allaby:1970jt,Antinucci:1972ib} (at the $\sqrt{s}=53$~GeV of the ISR, for example, DPMJET-III underestimates the $\bar p$ multiplicity by a factor of $\sim0.7$, while at $\sqrt{s}=6$~GeV~\cite{Allaby:1970jt} it overestimates it by a factor of $\sim2$).

The end result of this exercise is that the CR $\bar d$ flux prediction of~\cite{Ibarra:2013qt} is lower than that of~\cite{Duperray:2005si} by a factor of $\sim3$ (where we recall that the $\bar d$ fit of~\cite{Duperray:2005si} already underestimates the ISR $pp\to\bar d$ data~\cite{1973PhLB...46..265A,Henning:1977mt} by a factor of $\sim2$); and lower than our prediction by a factor of 10.

\underline{Herms et al~\cite{Herms:2016vop}} adopted their $\bar d$ production from Ref.~\cite{Ibarra:2013qt}, discussed above. However, for $\ah$ production, a different set of cross sections is used. We do not enter a detailed comparison to their results. However, we note that following~\cite{Carlson:2014ssa} (that focused on a dark matter source for $\ah$), the possibility is entertained that different coalescence momenta could apply to $\bar d$ and $\ah$ production, leading to a potential factor of $\gtrsim10$ increase in $B_3$ compared to what would naively be deduced by using the same $p_c$; this enhancement is compatible with what we suggest here as our baseline hypothesis.

A general comment should be added in comparing results based on event-by-event Monte-Carlo (MC) generators, as done in~\cite{Cirelli:2014qia,Ibarra:2012cc,Ibarra:2013qt,Herms:2016vop}, to the semi-analytic Eq.~(\ref{eq:coal}) that we used in the bulk of our analysis. As discussed, for example, in~\cite{Ibarra:2013qt}, the MC calculation takes into account correlations in the hard process that could modify the coalescence yield near threshold by a factor of  order unity. 
We can quantify the effect by comparing an event-by-event PYHTIA calculation to a calculation based on Eq.~(\ref{eq:coal}), where the underlying $\bar p$ cross section is taken consistently from the same PYTHIA tune. 

\begin{figure}[t]
\begin{center}
\includegraphics[scale=0.65]{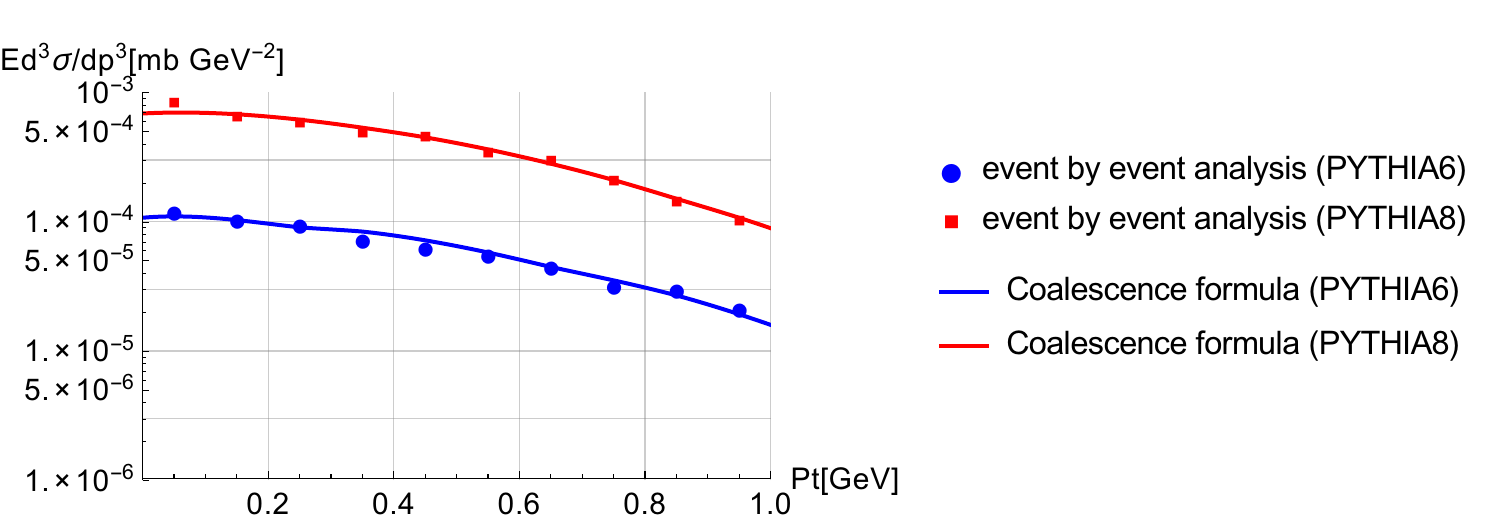}
\includegraphics[scale=0.65]{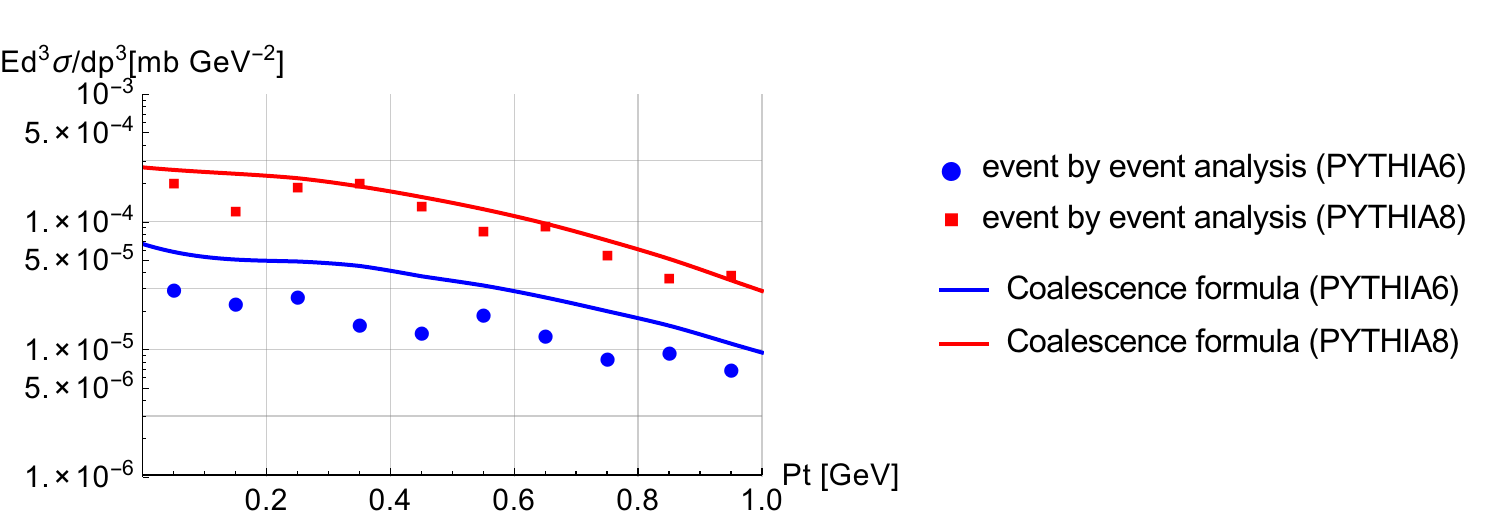}
\caption{Comparison of $pp\to\bar d$ production cross section, derived by the event-by-event method and by the analytic coalescence formula Eq.~(\ref{eq:coal}) using an underlying $pp\to\bar p$ cross section from the same PYTHIA tune. We use the PYHTIA 6 and 8 versions of Fig.~\ref{fig:pythia}. Top panel: $\sqrt{s}=53$~GeV. Bottom panel: $\sqrt{s}=20$~GeV. Both panels use $p_c=160$~MeV.}
\label{fig:pythia2}
\end{center}
\end{figure}

Our results are shown in Fig.~\ref{fig:pythia2}, focusing on $pp\to\bar d$ production. The top (bottom) panel correspond to $\sqrt{s}=53$~GeV (20~GeV), respectively. Above threshold -- in the top panel -- the MC calculation gives identical results to Eq.~(\ref{eq:coal}), for both PYTHIA tunes. At lower $\sqrt{s}$, in the bottom panel, the event-by-event PYTHIA 6 calculation falls below Eq.~(\ref{eq:coal}) by about a factor of 2, while the PYHTIA 8 event-by-event calculation is consistent with Eq.~(\ref{eq:coal}) to about 50\%. We conclude that the event-by-event calculation method is in reasonable agreement with Eq.~(\ref{eq:coal}), when both methods are based on the same underlying $pp\to\bar p$ cross sections. \\

Finally, Refs.~\cite{Chardonnet:1997dv,Duperray:2005si,Cirelli:2014qia,Ibarra:2013qt,Herms:2016vop} treated CR propagation within specific models, including the leaky-box and homogeneous thin disc+halo diffusion models, while we simply used the secondary relation Eq.~(\ref{eq:sec}). As demonstrated by the $\bar p$ data in Figs.~\ref{fig:flux}-\ref{fig:pppb}, apart from some potential $\mathcal{O}(10\%)$ effects that are unimportant in comparison to the particle physics cross section uncertainties, details of propagation are irrelevant to the calculation as long as we keep to the relativistic regime. At low rigidity $\R\lesssim5$~GV and correspondingly low energies, complications due to the details of propagation in the mildly non-relativistic regime; solar modulation; energy-dependent fragmentation cross sections entering the grammage analysis; etc., render the analysis complicated and model-dependent. For that reason throughout the paper we restricted ourselves to $\R>5$~GV.\\

\noindent
\mysection{Appendix C: Phase space correction}\label{sec:ph}
\noindent
Following~\cite{Duperray:2002pj,Duperray:2003tv}, we insert a phase space threshold correction $R(x)$ in Eq.~(\ref{eq:coal}), given by the ratio
\be R(x)&=&\frac{\Phi_N(x,m_p)}{\Phi_N(x,0)},\ee
where 
\be
\Phi_N(x,m)&=&\left[\Pi_{i=1}^N\int\frac{d^3p_i}{(2\pi)^32E_i}\right](2\pi)^4\delta^{(4)}\left(\sum_ip_i-P\right)\no\\
\ee
with $P=\left(x,\vec 0\right)$  
is the phase space of the $N=2+A$ mass $m$ nucleons of available CME $x$ that are minimally produced along with an mass number A anti-nucleus. 

\begin{figure}[t]
\begin{center}
\includegraphics[scale=0.8]{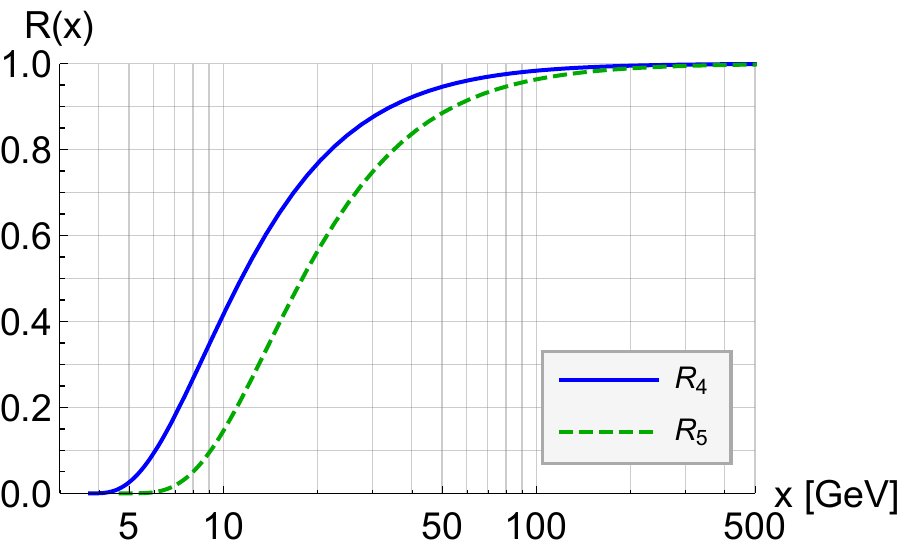}
\caption{$R(x)$. For $\bar d$ ($\ah$) calculations we use $R_4$ ($R_5$).}
\label{fig:R}
\end{center}
\end{figure}

Our results for $R(x)$ obtained by direct integration differ from the results in~\cite{Duperray:2002pj,Duperray:2003tv}. For reference we plot our result for $R(x)$ in Fig.~\ref{fig:R}, that can be directly compared to plots in~\cite{Duperray:2002pj,Duperray:2003tv}. To judge the numerical importance of the correction factor $R(x)$ to the analysis of laboratory data, note that typical kinematics in~\cite{BOZZOLI1978317,BUSSIERE19801} imply $x\sim15-20$~GeV, while the data in~\cite{Abramov:1986ti} corresponds to $x\sim9$~GeV.
\\

\noindent
\mysection{Appendix D: The $\bar pp\to\ah$ source}\label{sec:pbp}
\noindent
In the body of the paper we considered $\bar d$ and $\ah$ production in pp scattering. However, given the known CR $\bar p$ flux, some contribution to the CR $\bar d$ and $\ah$ flux should come from $\bar pp$ collisions. This source could in principle be important, despite the low $\bar p$ flux, if the production cross section of anti-nuclei is higher by a factor of $10^4$ or so compared to the pp cross section.

Light nuclei production in $\bar pp$ scattering was studied in Ref.~\cite{Alexopoulos:2000jk} at $\sqrt{s}=1.8$~TeV. 
We estimate that the cross section required for $\bar pp\to\at$ to produce an observable ratio, $\ah/\bar p\sim10^{-5}$, is on the order of $\left(\frac{d\sigma_{\bar pp\to\at}}{dy}\right)_{y=0}\sim1~\mu$b. The result of~\cite{Alexopoulos:2000jk} was $\left(\frac{d\sigma_{\bar pp\to t}}{dy}\right)_{y=0}\approx0.8$~$\mu$b; on general grounds, the process should satisfy $\sigma_{\bar pp\to\at}=\sigma_{\bar pp\to t}$. This could suggest that the contribution of $\bar pp$ collisions to the CR $\ah$ flux is significant. However, the experimental yield for d production reported in~\cite{Alexopoulos:2000jk} was surprisingly close to the t yield, with $\sigma_{\bar pp\to d}/\sigma_{\bar pp\to t}\sim3$. This is to be contrasted with an expected $\mathcal{O}(100)$ hierarchy between the $A=2$ and $A=3$ cross sections, a cause for concern that some systematic may be at play. For that reason we do not analyze in detail the $\bar pp\to t$ data of~\cite{Alexopoulos:2000jk}. Clarifying the validity of the $\left(\frac{d\sigma_{\bar pp\to\at}}{dy}\right)_{y=0}$ result of~\cite{Alexopoulos:2000jk}, through careful examination of possible systematics by the experimental collaboration, is highly motivated.\\

\noindent
\mysection{Appendix E: Input cross sections for grammage calculation}
\noindent
Our calculation of secondary CR anti-nuclei relies on experimental data of other secondary and primary CRs, notably the B/C and C/O flux ratios, in order to calibrate out the effect of CR propagation via the CR grammage $\X$. In turn, the derivation of $\X$ via Eqs.~(\ref{eq:Q}-\ref{eq:Xbc}) requires knowledge of decayed nuclear fragmentation cross sections. (For a recent collection of experimental references, see~\cite{Tomassetti:2015nha}.) The fragmentation cross section data is typically specified only at low energies, $\lesssim2-4$~GeV/nuc; we extrapolate this information to high energy assuming that the cross sections remain constant. The experimental uncertainty on the most relevant cross sections -- the reactions $^{12}$C$\to$$^{11}$B, $^{16}$O$\to$$^{11}$B -- is of order 20\%. The projectile specie included in the calculation and the benchmark cross section values we use are listed in Tab.~\ref{tab:fragXS}, and refer to fragmentation on hydrogen target. These  values are based on data summarised in Ref.~\cite{0954-3899-28-6-304,0954-3899-26-8-306,PhysRevC.28.1602,PhysRev.119.316,0004-637X-508-2-949,0004-637X-501-2-911,1993ICRC....2..187K}.
We extend the result to account for He in the ISM, assuming number density $n_{ISM}=0.9n_H+0.1n_{He}$ and using the formula in Ref.~\cite{Ferrando:1988tw}. 
For the total inelastic cross section of B, we use the formula in Ref.~\cite{Ferrando:1988tw}.
\begin{table}[htb]
\caption{Benchmark decayed fragmentation cross sections.}
\begin{tabular}{|c|c|c|c|}\hline
reaction&benchmark cross section (mb)\\ \hline
$^{12}$C$\to$$^{11}$B&55\\ \hline
$^{12}$C$\to$$^{10}$B&14\\ \hline
$^{16}$O$\to$$^{11}$B&25\\ \hline
$^{16}$O$\to$$^{10}$B&9\\ \hline
$^{14}$N$\to$$^{11}$B&30\\ \hline
$^{14}$N$\to$$^{10}$B&9\\ \hline
$^{20}$Ne$\to$$^{11}$B&14\\ \hline
$^{20}$Ne$\to$$^{10}$B&2\\ \hline
$^{24}$Mg$\to$$^{11}$B+$^{10}$B&15\\ \hline
\end{tabular}
\label{tab:fragXS}
\end{table}

\end{appendix}

\end{document}